 \documentclass[final,3p,times]{elsarticle_arxiv}

\usepackage{amssymb,amsmath}
\usepackage{mathtools}
\usepackage{longtable}
\usepackage{booktabs}
\usepackage[overload]{empheq}
\usepackage{color}

\usepackage{lineno}

\usepackage{graphicx}
\usepackage{lscape}
\usepackage{caption}
\usepackage{subcaption}
\newcommand*\rfrac[2]{{}^{#1}\!/_{#2}}
\newcommand{\CY}{\ensuremath{c_{y}}~}
\newcommand{\CYc}{\ensuremath{c_{y}}}
\newcommand{\BCYc}{\ensuremath{C_{Y}}}

\newcommand*{\review}{\textcolor{black}}

\journal{Mechanical Systems and Signal Processing DOI: http://dx.doi.org/10.1016/j.ymssp.2017.04.048}

\begin{document}

\begin{frontmatter}



\title{Nonlinear state-space modelling of the kinematics of an oscillating circular cylinder in a fluid flow}


\author[label1,label2]{J. Decuyper}
\ead{jan.decuyper@vub.ac.be}
\author[label2]{T. De Troyer}
\author[label2]{M. C. Runacres}
\author[label1]{K. Tiels}
\author[label1]{J. Schoukens}

\address[label1]{Vrije Universiteit Brussel (VUB), Department of Fundamental Electricity and Instrumentation (ELEC), Pleinlaan 2, 1050 Brussels, Belgium.}
\address[label2]{Vrije Universiteit Brussel (VUB), Department of Engineering Technology (INDI), Pleinlaan 2, 1050 Brussels, Belgium.}

\begin{abstract}

The flow-induced vibration of bluff bodies is an important problem \review{of many} marine, civil, or mechanical engineers. In the design phase of such structures, it is vital to obtain good predictions of the fluid forces acting on the structure. Current methods rely on computational fluid dynamic simulations (CFD), with a too high computational cost to be effectively used in the design phase or for control applications. Alternative methods use heuristic mathematical models of the fluid forces, but these lack the accuracy (they often assume the system to be linear) or flexibility to be useful over a wide operating range.

In this work we show that it is possible to build an accurate, flexible and low-computational-cost mathematical model using nonlinear system identification techniques. This model is data driven: it is trained over a user-defined region of interest using data obtained from experiments or simulations, or both. Here we use a Van der Pol oscillator as well as CFD simulations of an oscillating circular cylinder to generate the training data. Then a discrete-time polynomial nonlinear state-space model is fit to the data. This model relates the oscillation of the cylinder to the force that the fluid exerts on the cylinder. The model is finally validated over a wide range of oscillation frequencies and amplitudes, both inside and outside the so-called lock-in region. We show that forces simulated by the model are in good agreement with the data obtained from CFD.
\end{abstract}


\end{frontmatter}


\section{Introduction}
\label{intro}
The kinematics of oscillating bluff bodies in a fluid flow have been a hot research topic for decades. Most important are the potentially harmful vortex-induced vibrations (VIV) where the structure is excited by alternating vortex shedding. Classical examples are the vibration of chimney stacks exposed to wind, pipelines on the sea-bed excited by the ocean currents or tubes in heat exchangers \cite{williamson1996}.

For this kind of fluid-structure interactions, typically an analytical solution cannot be found. However, accurate predictions of the kinematics and the resulting dynamics are vital for instance in the design process, in monitoring applications or for control. To obtain predictions with high fidelity, so far only two possibilities existed: solving the Navier-Stokes equation via computation fluid dynamic (CFD) simulations or performing experiments. Both approaches are time-consuming, expensive, and require dedicated lab or computing facilities. These drawbacks make the current methods too expensive for many intended applications, where only limited time and resources might be available. 

To tackle the shortcomings of CFD and experiments, low-order models of VIV have been developed \cite{gabbai2004}. Most models are heuristic, with only a limited range of applicability. Different types exist, the most noteworthy can be classified as force decomposition models \cite{sarpkaya2004}, single degree-of-freedom (sdof) models  \cite{simiu1986} or wake oscillator models, where the wake oscillation is generally formulated as a Van der Pol equation \cite{parkinson1974,hartlen1970,facchinetti2004}.

Although capable of qualitatively describing the characteristic behaviour of VIV, an efficient and powerful model, flexible enough to span a wide domain in parameter space with a single set of parameters has yet to be obtained. To identify such a model, system identification can be a very powerful tool. In recent years, the application of linear system identification techniques has already shown some promising results, e.g.\ the best linear approximation in least squares sense \cite{runacres2013} or building an input-output relationship using delayed values of the input and the output (ARX model) \cite{zhang2015}.

Nevertheless, the challenges in modelling the system at hand are substantial. Fluid-structure interactions are, for one, inherently nonlinear \cite{bishop1963}. \review{This is strongly pronounced by the fact that VIV is a selfexcited yet self-limited oscillation, resulting in a stable limit cycle \cite{pikovsky2001}}. In addition, it has been shown that the vortex shedding behaviour is hysteretic \cite{govardhan2000}. Nonlinear modelling of VIV is currently a very active research topic. In \cite{Wang2003}, the auto-regressive moving averaging (ARMA) technique is combined with the modal analysis method. A nonlinear force representation is obtained by including higher harmonic terms resulting from the fluid-structure coupling. The coefficients in the model are determined to provide a best fit to the measured time series in maximum-likelihood sense. Reasonable predictions in terms of maximum amplitude are found, although only a limited amount of validation cases were investigated. In recent work \cite{winter2016}, local linear models are used as a surrogate model for aerodynamic loads on wings. A general framework is however not yet in place when it comes to nonlinear identification of complex dynamical systems. A lot can still be gained in terms of quality and applicability of the model but also in terms of insight into the nonlinear behaviour of VIV.

In this work, we propose a novel approach to model flow problems that applies state of the art nonlinear identification techniques. The powerful framework of nonlinear state-space models is explored and put to the test. Nonlinear state-space models have proven to be very rich and flexible, providing solutions to a variety of nonlinear problems also including hysteresis \cite{widanage2011,noel2017}. 

The methodology of this work can be summarised as follows:
\begin{enumerate}
\item Use established methods, CFD simulations in this case, although the proposed method is equally suitable for experiments, to study the fluid forces within a user defined domain of parameter space. Then capture these observations under the form of time series.
\item Identify a nonlinear black-box model that is able to reproduce the observed phenomenon.
\item Exploit the benefits of the obtained model to simulate, at very low cost, a variety of possible regimes of the system.
\end{enumerate}  

The lay-out of the paper is as follows: in Section \ref{sim_set_up}, the simulation set-up is introduced. Section \ref{black-box} describes the discrete polynomial nonlinear state-space model structure and how such a model is identified. Section \ref{application} consists of two parts. In the first part, the method is applied to a nonlinear oscillator of the Van der Pol type. In the second part the kinematics of an oscillating cylinder is addressed. Section \ref{conclusion} concludes the paper.

\section{Simulation set-up}
\label{sim_set_up}
\begin{figure}[h!]
\begin{center}
\includegraphics[width=\textwidth]{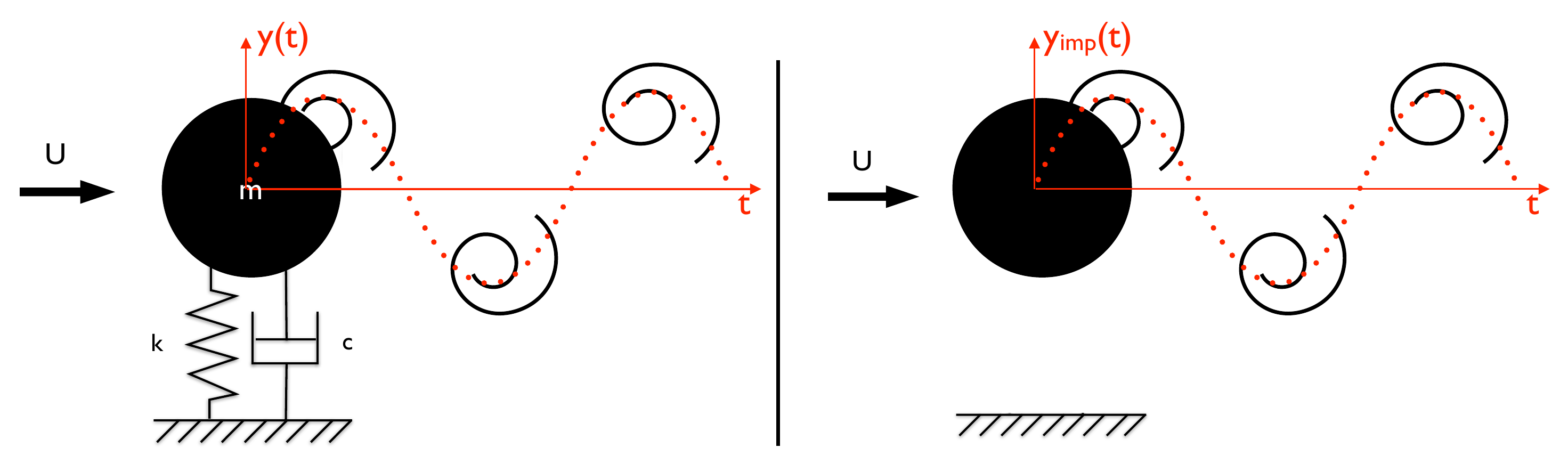}
\caption{Left: set-up of the self-excited vibration where $m$ indicates the mass of the cylinder, $k$ the stiffness, $c$ the damping coefficient and $y(t)$ the displacement of the cylinder. Right: The imposed oscillation set up, independent of structural parameters, with $y_{\text{imp}}$(t) the imposed displacement\review{, in the crosswise direction on the incoming flow}.}
\label{f:mass_spring_damper1}
\end{center}
\end{figure}

VIV are often studied on a 2D circular cylinder, suspended by a spring and a damper and restricted to move in the transverse direction, perpendicular to the incoming flow. Such a set-up is a self-excited oscillator, where the fluid forces generate the displacement. 
Common practice, however, 
is the use of imposed (or forced) oscillations \cite{govardhan2000,morse2009,bearman1984} to study the interaction between the flow field and the cylinder. With total control of the displacement one can explore a wide range of frequencies and amplitudes, contrary to the limited response behaviour observed when the mass, stiffness and the damping are fixed to certain values. The set-up of both the free and the imposed oscillations can be seen in Fig.\ \ref{f:mass_spring_damper1}.

\subsection{Input-output description of the system}
Since the displacement is imposed, this is logically considered the input to the system. To quantify the behaviour of the wake one could consider any of the classical fluid variables: velocity, vorticity or pressure in some points downstream. Although not all variables are equally informative \cite{decuyper2014}, they remain good candidates to build a model from. More obvious, from an application point of view, is a displacement-force relationship. Moreover, the fluid force on the cylinder can be seen as a combined fluid variable, concentrated in a single observable point, the centre of the cylinder. The system can thus be considered as a single input-single output system. To generalise the findings, both the input and the output are made dimensionless. The input is given by $A/D$ with $A$ the amplitude of the displacement and $D$ the diameter of the cylinder. The output is given by the force coefficient \CYc,
\begin{equation}
\label{e:cy}
c_y=\frac{F_y}{\rfrac{1}{2}\rho U^2D},
\end{equation}
where $F_y$ is the resultant of the forces acting on the cylinder in the $y$-direction, \review{ perpendicular onto the incoming flow, $U$ is the unperturbed flow speed, $\rho$ is the density of the fluid.}

\subsection{The autonomous oscillator \CY}
The frequency of natural vortex shedding, $f_\text{St}$, is given as Strouhal's relationship:
\begin{equation}
\label{e:fSt}
S=\frac{f_{\textup{St}}D}{U},
\end{equation}
where $D$ is the diameter and $U$ is the fluid velocity of the unperturbed flow \cite{strouhal1878}. $S$ is the so-called Strouhal number, which is a slowly varying function of the Reynolds number (Re) and has a value of about 0.16 around $\text{Re}=100$ \cite{blevins1990}.

Even in the case of a static cylinder, \CY will oscillate at $f_{\text{St}}$. In other words\review{, \CY behaves as an autonomous oscillator, tracking a stable limit cycle \cite{pikovsky2001}}. If however an excitation (imposed oscillation) of certain frequency and amplitude is applied, the frequency of vortex shedding can deviate from the natural vortex shedding frequency and synchronise to the excitation frequency. The region of parameter space for which this synchronisation phenomenon is observed is called the lock-in region \cite{blevins1990}. The lock-in region is of special interest since within this region the maximum values of the fluid forces are observed \cite{williamson1996}.

\subsection{CFD simulations of an oscillating circular cylinder}
\label{CFD}
Using CFD simulations of the flow around cylinders with imposed oscillations, time series of the fluid variables \review{are} obtained. \review{Broadband signals that span a frequency band around $f_{\text{St}}$ are chosen as excitation signals.} The simulations are performed using the open source CFD package OpenFOAM \cite{openFoam}. A transient solver called \emph{pimpleDyMFoam} with a variable time step was used to solve the \emph{Reynolds averaged Navier Stokes} (RANS) equations \review{for an incompressible flow} in a finite volume discretisation scheme \review{ \cite{islam2014,versteeg2007}}. \review{All simulations were performed at $\text{Re}=100$ to ensure laminar, predominantly 2D \cite{williamson1996}, vortex formation. The unperturbed velocity $U$ is set by the Strouhal relationship (Eq.\ \ref{e:fSt}), selecting $f_{\text{St}}=3$ Hz results in $U=1.7964$ m/s for $\text{S}=0.167$ (this is the uncorrected $S$, see below). The cylinder diameter $D=0.10$ m. Consequently the kinematic viscosity $\nu=1.7964\times10^{-3}\frac{\text{m}^2}{\text{s}}$. Since the flow is laminar, no turbulence model was used. The time step was adjusted by the solver, satisfying at all times the Courant condition, $\text{Cr}<0.5$ with Cr the Courant number. After transients have settled the time step generally converged to a value $O(2\times10^{-5})$ s. From these raw data, time series are extracted using linear interpolation at a fixed sampling frequency of $f_s=50$ Hz. Given that  the highest excited frequency is 4.5 Hz (see Section \ref{est_dataset}), output signals containing up to the fifth harmonic will be alias-free.} 

\review{Since the vortex shedding is only 2 dimensional at $\text{Re}=100$}, a 2D mesh with the cylinder positioned 10 $D$ from the inlet and centred in the $y$-direction was used. The computation domain is $40 D$ long ($L$) and $30 D$ high ($H$). The top and bottom of the domain are constrained by a slip boundary condition while on the cylinder surface itself, \review{the velocity is set equal to the grid velocity, which follows from the cylinder motion (no-slip condition)}. At the inlet, a uniform velocity profile is used. The outlet is conditioned with a zero velocity gradient. 

\review{Validation of the CFD models was carried out based on vortex shedding related quantities, which are tabulated in Table \ref{t:val_CFD}. Two meshes were evaluated, a fine mesh containing a total of 55362 hexahedra cells and a coarser one of 35172 hexahedra cells. The tested quantities showed to be independent from the mesh used with deviations well below 1 \%. All the data that were used for identification was produced using the finest mesh, although the coarser one would have sufficed given the accuracy when compared to literature.}

\review{For the aspect ratio of our computational domain ($H/D=30$), we can expect a small effect of blockage on the force coefficients and Strouhal number.  \cite{placzek2009,qu2013}.
To compensate for this blockage effect, the free stream velocity was correct from the formula derived in \cite{allen1944} and quoted in \cite{roshko1961,anagnostopoulos1994,anagnostopoulos1997},}
\begin{equation}
\label{e:U_corr}
\frac{U'}{U}=1+\frac{1}{4}c_d\left(\frac{D}{H}\right)+0.82\left(\frac{D}{H}\right)^2,
\end{equation}
\review{where $U$ and $U'$ are the real and the corrected free stream velocity correspondingly, $D$ is the diameter of the cylinder, $H$ is the height if the computational domain and $c_d$ is the uncorrected mean drag coefficient.}

\review{In \cite{anagnostopoulos1994} it was suggested that $c_d$ can be corrected from:}
\begin{equation}
\label{e:Cd_corr}
\frac{c'_d}{c_d} = 1-\frac{1}{2}c_d\left(\frac{D}{H}\right)-2.5\left(\frac{D}{H}\right)^2,
\end{equation}

\begin{figure}
\begin{center}
\begin{subfigure}[b]{0.45\textwidth}
\includegraphics[width=\textwidth]{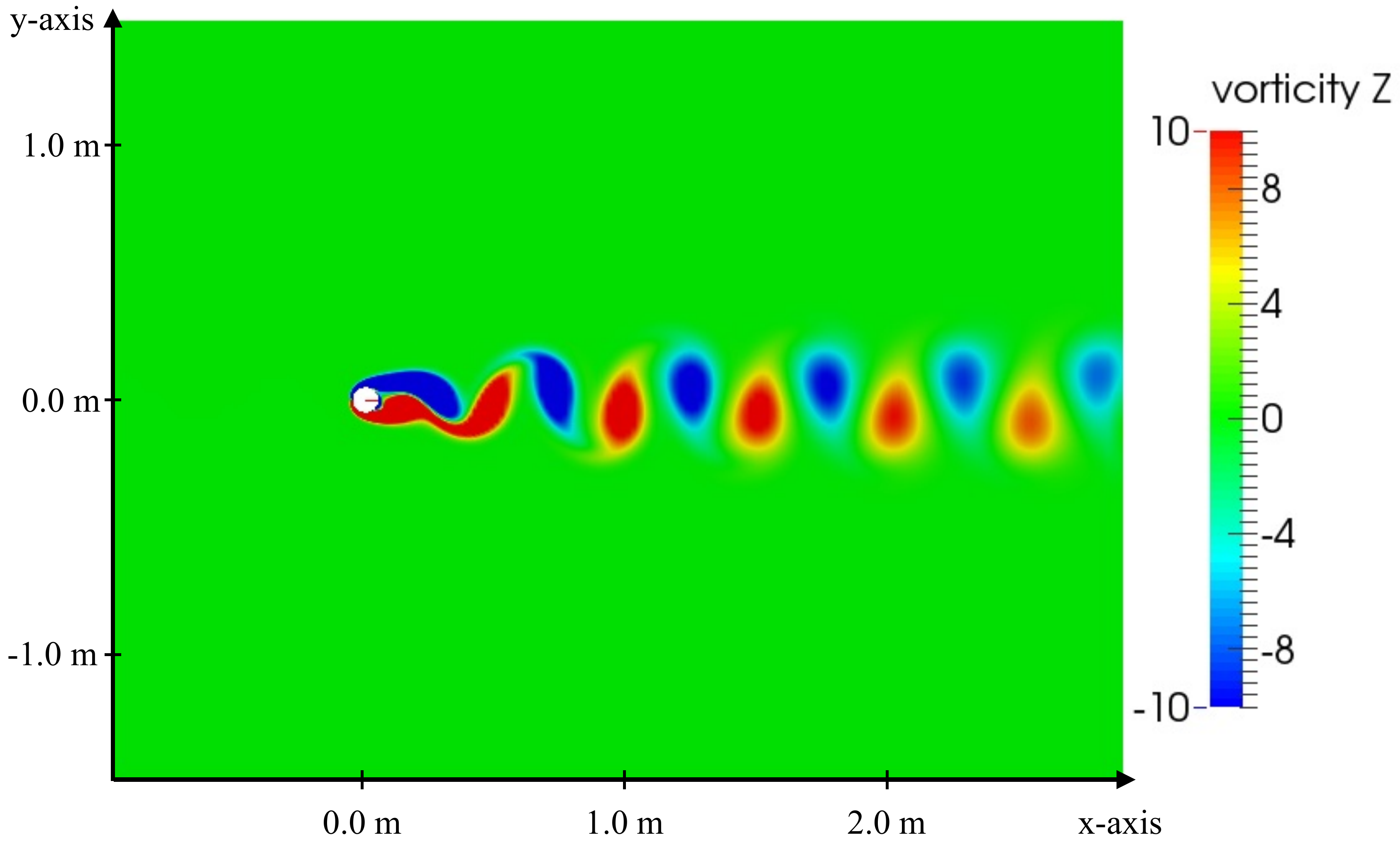}
\caption{}
\end{subfigure}
\begin{subfigure}[b]{0.45\textwidth}
\includegraphics[width=\textwidth]{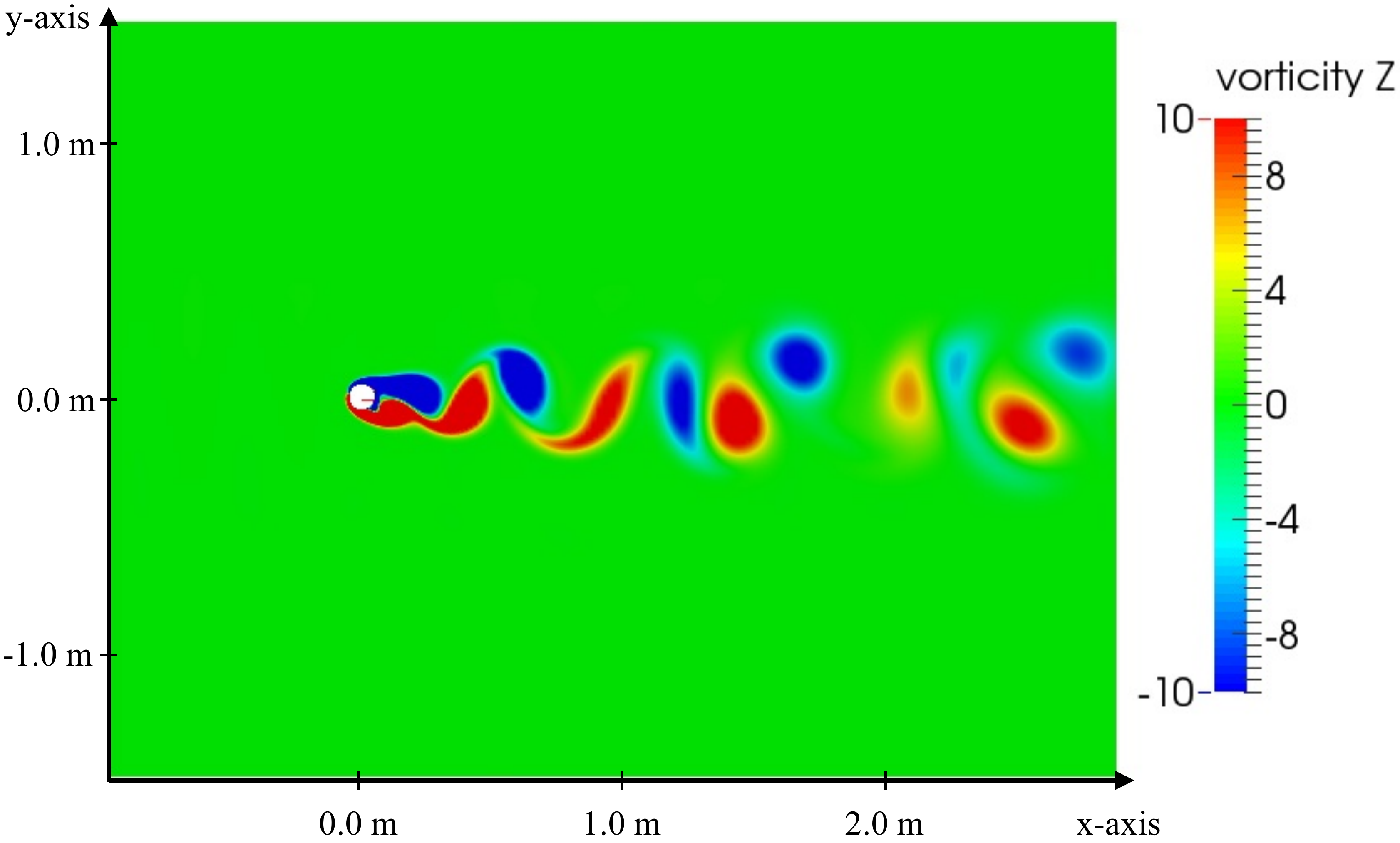}
\caption{}
\end{subfigure}
\begin{subfigure}[b]{0.45\textwidth}
\includegraphics[width=\textwidth]{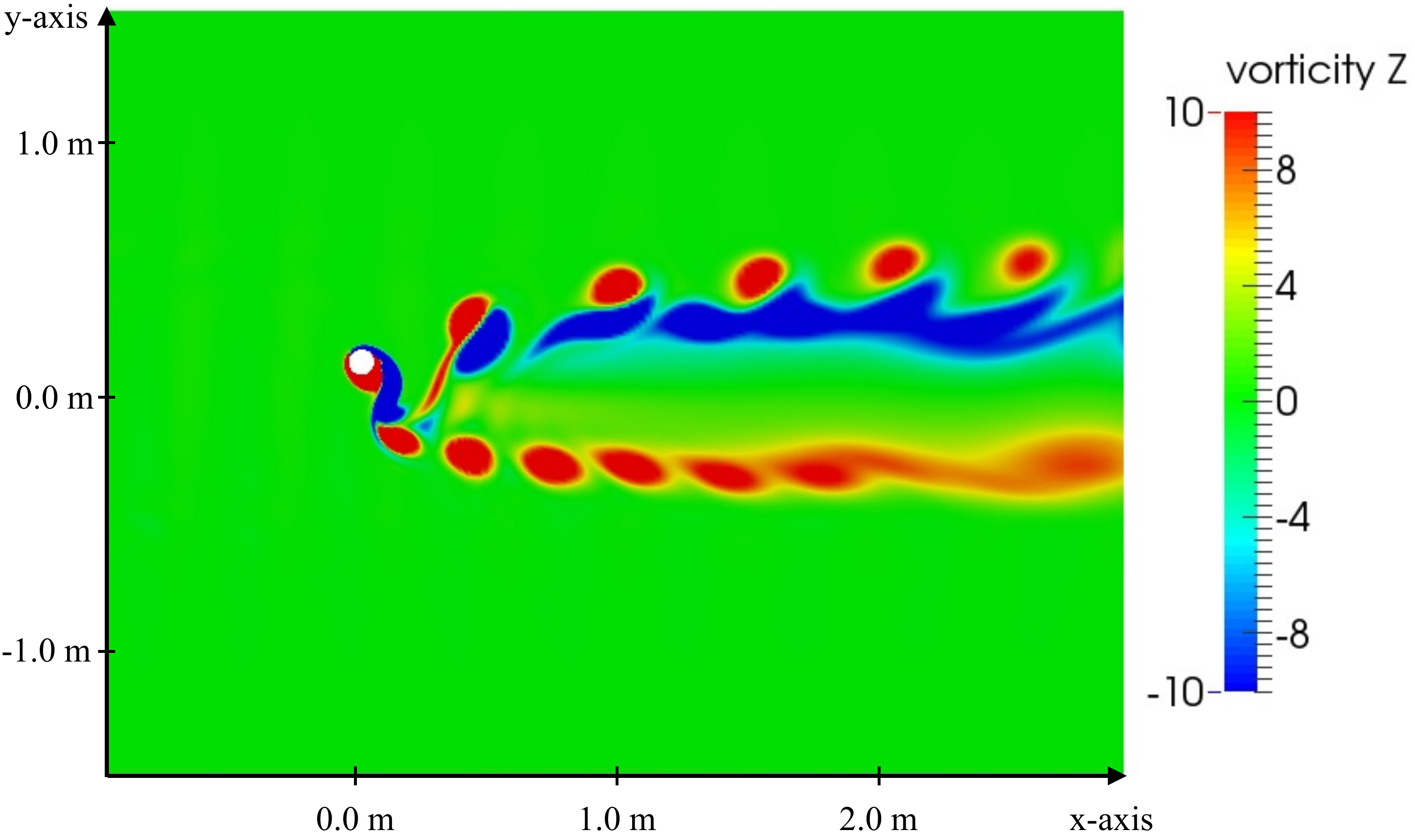}
\caption{}
\end{subfigure}
\begin{subfigure}[b]{0.45\textwidth}
\includegraphics[width=\textwidth]{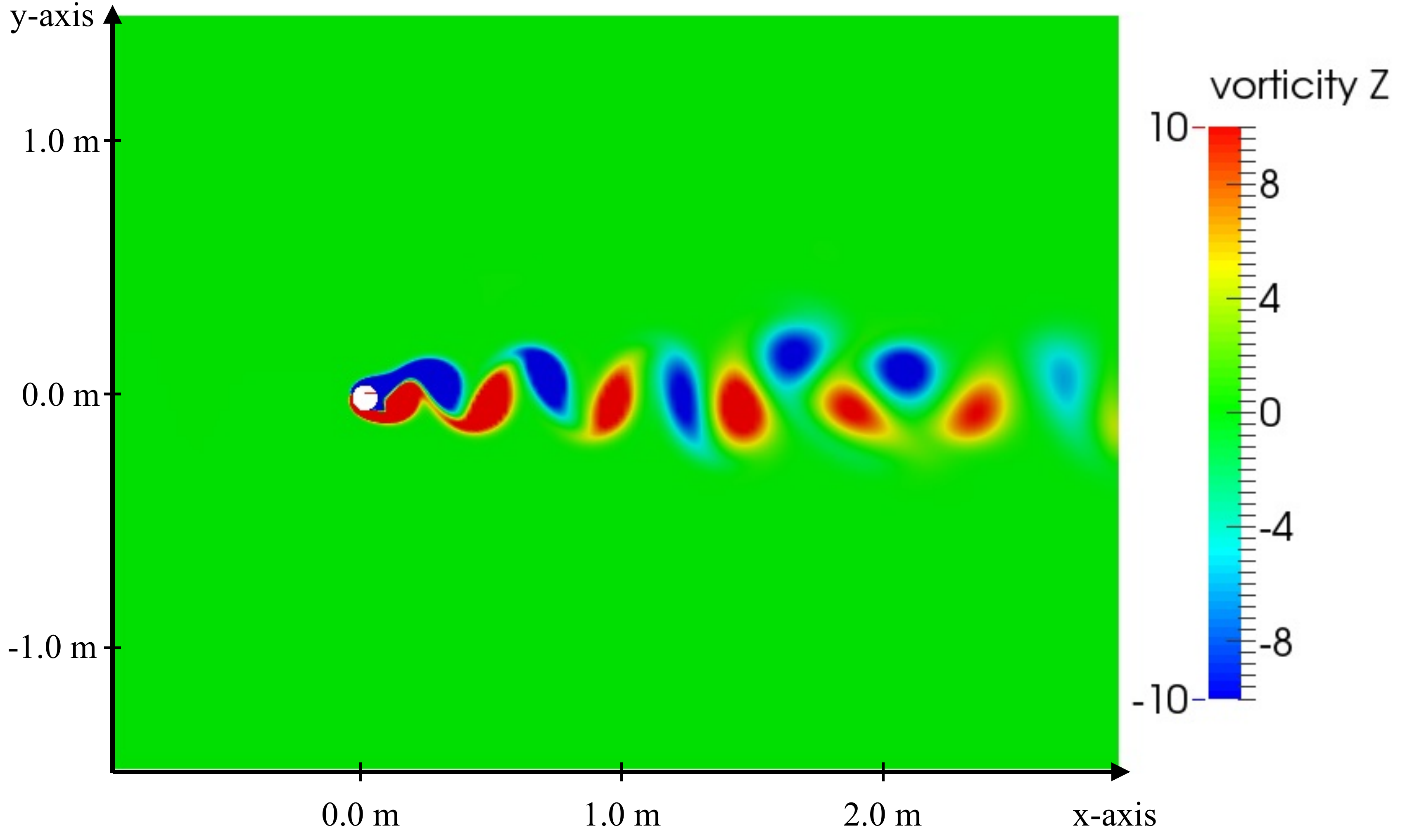}
\caption{}
\end{subfigure}
\caption{Image of the vortex formation in the wake behind a 2-dimensional circular cylinder at $\text{Re} = 100$. Figure a: stationary cylinder. Figure b: imposed sine sweep displacement at $A/D=0.20$. The shot is taken while sweeping through the lock-in region. Figure c: imposed single sine displacement of $\rfrac{f}{f_{\text{St}}}=1.5$ and $A/D=1.25$. The so-called P+S wake mode \cite{bearman2009}, where both a single vortex and a pair of opposite spinning vortices are shed in each period, is observed. Figure d: imposed random-phase multisine displacement with $f_0=0.1$ Hz and 45 excited harmonics at $A/D=0.3$ \review{(see Section \ref{init_BLA})}.}
\label{f:CFD_screenshots}
\end{center}
\end{figure}

\review{with $c_d$ the uncorrected drag coefficient and $c'_d$ the corrected one. It was however noted that the discrepancy would be very small if instead of utilising Eq.\ \ref{e:Cd_corr}, the drag coefficient would be calculated using the corrected free stream velocity $U'$. Using $U'$ also \CY and the Strouhal number can be corrected. Note that a correction in $U$ also implies a correction in Re. The true Reynolds number, given the present blockage ratio is approximately equal to 101.}
Fig.\ \ref{f:CFD_screenshots} illustrates the vortex formation in the wake behind an oscillating cylinder.

\begin{table}[h!]
\begin{center}
\begin{tabular}{| l | c  c  c |}
\cline{2-4}
 \multicolumn{1}{c|}{} & $c''_y$ & $c'_d$ & $S'$ \\
\cline{1-4}
Present study, fine mesh & 0.223 & 1.331 & 0.165\\
Present study, coarse mesh & 0.217 & 1.331 & 0.165 \\
\hline
\hline
Park et al., 1998 \cite{park1998} & 0.235 & 1.33 & 0.165\\
Kravchenko and Moin, 1998 \cite{kravchenko1998}& 0.222 & 1.32 & 1.64\\
Mittal, 2005 \cite{mittal2005}& 0.226 & 1.322 & 1.644\\
St\aa lberg et al., 2006 \cite{stalberg2006}& 0.233 & 1.32 & 0.166\\
Posdziech and Grundmann, 2007 \cite{posdziech2007}& 0.228 & 1.325 & 0.1644\\
Qu et al., 2013 \cite{qu2013} & 0.225 & 1.319 & 0.1648\\
\hline
\end{tabular}
\end{center}
\caption{Validation results on stationary cylinder simulations at Re$=100$ where $c''_y$ denotes the rms value of the corrected force coefficient in the $y$-direction (perpendicular on the oncoming flow), $c'_d$ is the mean value of the corrected drag coefficient and $S'$ is the corrected Strouhal number in the case of the present study and the reported values from literature found in \cite{qu2013}.}
\label{t:val_CFD}
\end{table}

%
%

\section{Black-box identification using a polynomial nonlinear state-space model}
\label{black-box}

System identification generally requires a four-step approach. \review{First of all,} it is essential to collect data of the user-defined problem (Section \ref{excitation}). To that end, an experiment, tailored to the problem at hand, is designed. Secondly a model structure is selected (Section \ref{model_structure}). The model structure has to be flexible or generic enough to be able to capture the observed behaviour of the system. \review{The overall goal will be to match} the model as well as possible to the data. \review{A part of this third step will be to evaluate the fit of the model. Therefore the notion of cost is introduced (Section \ref{costFunction})}. \review{To obtain a model estimate, the objective will be to minimise the cost, given the data and the selected model structure (Section \ref{IDPNLSS})}. Finally, the model is validated on a new dataset, not used during the identification, to avoid an over-dependence on stochastic contributions (Section \ref{valModel})\review{. This step will also require estimates of the initial states and input (Section \ref{Estx0}).}

\subsection{Experiment design}
\label{excitation}
The fact that it is unlikely that the true model is found implies that the identified model becomes dependent on the provided data. It is therefore crucial that the coefficients of the model are tuned on data which span the operating domain as well as possible, \review{in this way} the errors are pushed into regions that are of little or no interest.
When selecting the excitation signals of the training data, not only the frequency band and the amplitude, but also the type of signal must be taken into account. The intended operating conditions, richness and compactness of the data are all important factors. The selection of an adequate training set for identification of \CY is discussed in Section \ref{est_dataset}.

\subsection{Model structure}
\label{model_structure}

To be able to work with a data-based black-box modelling approach, the model structure has to be flexible enough to cover the problem of interest. The Polynomial Nonlinear State-Space (PNLSS) structure \cite{paduart2010} is able to provide such flexibility. An additional advantage of the state-space representation is that it is very well suited to incorporate multiple-inputs and multiple-outputs (MIMO). For now, only single-input single-output (SISO) of a 2D structure is \review{modelled}. \review{In future work it might be considered to exploit the MIMO abilities of the structure to relate the displacement of multiple spanwise cross sections of a 3D structure to their corresponding force coefficients. It should be noted that pressure fluctuations at certain cross sections may nonlinearly effect others. In recent work \cite{winter2016_neurofuzzy,skujins2014} this aspect has already been addressed using the aerodynamic strip theory. MIMO nonlinear state-space models could form an interesting alternative.} 

The nonlinear state-space model structure is defined as a classical linear state-space model where the state and the output equation are extended with some nonlinear functions $f$ and $g$ Eq.~\eqref{e:PNLSS1}. The model is built in discrete time. To maintain a nomenclature suited to the model to be built, $\CY$is used as output variable and $y$ is the input, which corresponds to the displacement of the cylinder in the $y$-direction.
\begin{subequations} \label{e:PNLSS1}
    \begin{empheq}[left={\empheqlbrace\,}]{align}
      & x(t+1)=\textbf{A}x(t)+\textbf{B}y(t)+f(x(t),y(t)) \\ 
      & \CYc(t)=\textbf{C}x(t)+\textbf{D}y(t)+g(x(t),y(t))
    \end{empheq}
\end{subequations}
When \review{the linear part is of} $n_x$-th order ($n_x$ state variables), the state vector $x(t)$ $\in \mathbb{R}^{n_x}$, the input vector  $y(t) \in \mathbb{R}^{n_y}$ with $n_y$ equal to the number of inputs and $\CYc(t) \in \mathbb{R}^{n_{Cy}}$, \review{which is} the vector of all the $n_{Cy}$ outputs. From hereon only the SISO case is considered where $n_{c_y}=n_y=1$. Correspondingly, the dimensions for the matrices of coefficients become: $\textbf{A} \in \mathbb{R}^{n_x\times n_x}$, $\textbf{B} \in \mathbb{R}^{n_x\times 1}$, $\textbf{C} \in \mathbb{R}^{1\times n_x}$, $\textbf{D} \in \mathbb{R}^{1\times 1}$.

In the PNLSS case, the nonlinear functions $f$ and $g$ are expanded in basis functions, leading to Eq.~\eqref{e:PNLSS2}\review{:}
\begin{subequations} \label{e:PNLSS2}
    \begin{empheq}[left={\empheqlbrace\,}]{align}
      & x(t+1)=\textbf{A}x(t)+\textbf{B}y(t)+\textbf{E}\zeta(t) \label{e:PNLSS2_a} \\ 
      & \CYc(t)=\textbf{C}x(t)+\textbf{D}y(t)+\textbf{F}\eta(t) \label{e:PNLSS2_b},
    \end{empheq}
\end{subequations}
where vectors $\zeta(t) \in \mathbb{R}^{n_\zeta}$ and $\eta(t) \in \mathbb{R}^{n_\eta}$ contain nonlinear monomials in $x(t)$ and $y(t)$. The monomials range from degree 0 up to a chosen degree $p$ for the state equation Eq.~\eqref{e:PNLSS2_a} and $q$ for the output equation Eq.~\eqref{e:PNLSS2_b}. All degrees within this range are included, except for degree 1, since these contributions are already covered by the linear part of the structure. The monomials $\zeta(t)$ and $\eta(t)$ are formed by all possible products of the input and the \review{state} variables raised to a chosen degree, e.g.\ for an element in  $\zeta(t)$:
\begin{equation}
\label{e:PNLSS3}
\zeta_{k,l_1,...,l_{nx}}(t)=y^k(t)\prod_{i=1}^{n_x} x_i^{l_i}(t)
\end{equation}
with the total degree of the monomial satisfying the condition: $k+\sum_{i=1}^{n_x} l_i \in \left\{0,2,3,...,p \right\}$ with $[k,l_i] \in \mathbb{N}$ \cite{paduart2010}. \review{Therefore} \textbf{E} $\in \mathbb{R}^{n_x\times n_{\zeta}}$ and \textbf{F} $\in \mathbb{R}^{n_x\times n_{\eta}}$, with $n_{\zeta}$ and  $n_{\eta}$ the number of combinatorial products between the states and the input, satisfying the condition on the total degree of the monomial up and till $p$ for \textbf{E} and $q$ for \textbf{F}.

\subsection{The cost function}
\label{costFunction}
In this work, the cost is defined as the squared error between the output of the model and the training data \CYc. A time-domain cost function was preferred to facilitate the use of signals of different types and with different frequency content. Since the cost is dominated by model errors and not by noise, no noise weighting is applied. The least-squares cost function is given by:
\begin{equation}
\label{VLS}
V(\theta) =  ||e(t,\theta)||^2 \quad \text{with} \quad e(t,\theta) = c_{y\text{,CFD}}(t)-c_{y\text{,mod}}(t,\theta),
\end{equation}
where $\theta$ is the vector of parameters to be optimised. These parameters are all the coefficients in the \textbf{A,B,C,D,E} and \textbf{F} matrices: 
\begin{equation}
\label{theta}
\theta = [\text{vec(}\textbf{A}); \text{vec(}\textbf{B}); \text{vec(}\textbf{C}), \text{vec(}\textbf{D}); \text{vec(}\textbf{E}); \text{vec(}\textbf{F})],
\end{equation}
here vec denotes the operator which stacks all the elements of a matrix in a single column vector. $c_{y\text{,CFD}}$ denotes the \review{training} data while $c_{y\text{,mod}}$ is the simulated output of the model.
\subsection{Minimising the cost function} 
\label{IDPNLSS}

The procedure of identifying a PNLSS model consists of 3 major steps \cite{paduart2010}:
\begin{enumerate}
\item First, the best nonparametric linear approximation \review{(FRF) of the system} is estimated. The best linear approximation, denoted $\hat{G}_{\small{\textup{BLA}}}$, of a nonlinear system is defined as the model $G$, belonging to the set of all linear models $\mathcal{G}$ such that
\begin{equation}
\label{GBLAARGMIN}
\hat{G}_{\small{\textup{BLA}}} = \underset{G \in \mathcal{G}}{\operatorname{arg~min}}~\mathbb{E}(|C_Y(j\omega) - G(j\omega)Y(j\omega)|^2),
\end{equation}
where \BCYc$(j\omega)$ and $Y(j\omega)$, denote the discrete Fourier transforms of \CYc, and $y$ respectively. 
\item Next, linear identification techniques are used to obtain a linear parametric estimate of the model, \review{$\hat{G}_{\textup{BLA,par}}(\theta_{\text{lin}})$ (see Eq.\ \ref{e:theta_lin} for a definition of $\theta_{\text{lin}}$).}
\item Finally, all coefficients are further tuned to data via nonlinear optimisation of the cost function Eq.~\eqref{VLS}. As initial values for the $\textbf{A,B,C}$ and $\textbf{D}$ matrices the parametric BLA is used, $\textbf{E}$ and $\textbf{F}$ are set to zero.
\end{enumerate}

\subsubsection{Initialisation of the nonlinear model via the best linear approximation}
\label{init_BLA}
The best linear approximation, as defined in Eq.~\eqref{GBLAARGMIN}, is used to initialise the full nonlinear model. To obtain the best linear approximation, $\hat{G}_{\text{BLA}}$, random-phase multisine realisations are used \cite{dhaene2005}. The input signal, $y(t)$, corresponds to a sum of $N$ harmonically related sines with a user-defined amplitude spectrum. The phases $\phi_n$ are random variables from a uniform distribution between 0 and $2\pi$.
\begin{equation}
\label{e:MS1}
y(t) = \frac{1}{\sqrt{N}}\sum_{n=1}^{N} A_n\sin \left(2\pi n f_0 t+ \phi_n \right) \textup{ with } \phi_n \sim U[0,2\pi[
\end{equation}
In general, $M$ different random phase multisines are applied and $P$ number of periods of the input and the output signal are measured. Per experiment $m$ and period $p$ the discrete Fourier \review{transform} (DFT) of the input signal ($Y(k)^{[m,p]} \in \mathbb{C}$) and of the output signal ($\BCYc(k)^{[m,p]} \in \mathbb{C}$) are computed. The spectra are then averaged over the periods. At frequency line $k$ we get:
\begin{equation}
\label{e:average_spec}
\begin{multlined}
\begin{aligned}
 \hat{Y}(k)^{[m]} = \frac{1}{P}\sum_{p = 1}^{P}Y(k)^{[m,p]} & \\
 \hat{\BCYc}(k)^{[m]} = \frac{1}{P}\sum_{p = 1}^{P}\BCYc(k)^{[m,p]}. &
 \end{aligned}
\end{multlined}
\end{equation}  
For every experiment $m$, the transfer function estimate, $\hat{G}(z_k)^{[m]} \in  \mathbb{C}$, with $z_k=e^{j\frac{2\pi f_sk}{N}}$, $f_s$ the sampling frequency and $N$ the number of frequency lines, then becomes:
\begin{equation}
\hat{G}(z_k)^{[m]} = \frac{\hat{\BCYc}(k)^{[m]}}{\hat{Y}(k)^{[m]}}.
\end{equation}
Finally, the transfer function estimates are averaged over the realisations to get the Best Linear Approximation:
\begin{equation}
\label{e:GBLA}
\hat{G}_{\small{\textup{BLA}}}(z_k) = \frac{1}{M} \sum_{m=1}^{M} \hat{G}(z_k)^{[m]}.
\end{equation}

An advantage of using such periodic excitations is that the influence of nonlinear distortions can be separated from measurement noise \cite{schoukens1998}. Variations \review{found over the $P$ periods of the measured transfer functions} are caused by measurement noise while variations over the \review{transfer functions of }$M$ experiments \review{(phase realisations)} are due to the combined effect of measurement noise and nonlinear distortions (see Section \ref{modelEst}). 

From this nonparametric estimate and an estimate of the sample variance ($\hat{\sigma}^2_{\textup{BLA}}$), a parametric linear state-space model, $\hat{G}_{\textup{BLA,par}}$, is estimated using the frequency domain identification toolbox (FDIDENT\footnote{http://vubirelec.be/knowledge/downloads}) in MATLAB. Here the nonlinear cost function is minimised in the frequency domain. \review{The cost function is weighted by the total variance which contains contributions of both the noise and the nonlinear distortions:}
\begin{equation}
\label{e:BLApar}
V(\theta_{\text{lin}})=\sum_{k=1}^N \frac{| \hat{G}_{\small{\textup{BLA}}}(z_k) - \hat{G}_{\small{\textup{BLA,par}}}(\textbf{A},\textbf{B},\textbf{C},\textbf{D},z_k)|^2} {\hat{\sigma}^2_{\textup{BLA}}}
\end{equation}
where the vector of all the linear coefficients is given by:
\begin{equation}
\label{e:theta_lin}
\theta_{\text{lin}} = [vec(\textbf{A}); vec(\textbf{B}); vec(\textbf{C}); vec(\textbf{D})].
\end{equation}
\subsubsection{Nonlinear optimisation of the full model}
\label{Optimisation_strategy}


The full nonlinear model is obtained by minimising the least-squares cost function Eq.~\eqref{VLS} with respect to $\theta$. The optimisation is performed using the Levenberg-Marquardt (LM) algorithm. It is well known that gradient descent based methods, such as LM, are very sensitive to the choice of initial values. Given the starting values, the optimisation can be seen as a descending path along the cost function. Unfortunately, the optimisation might easily end up in a poor local minimum, close to the starting point. To increase the chances of ending up in a sufficiently good local minimum, one would like to descend the cost function over multiple paths, in different descending directions. A way to come to different descending paths, from a certain starting point, is by calculating the gradient based on a subset of the parameters, \review{as} opposed to all parameters simultaneously. Keeping a number of parameters constant during optimisation reduces the degree of freedom, resulting in a different path. 

\begin{figure}[h]
\begin{center}
\includegraphics[scale=0.4]{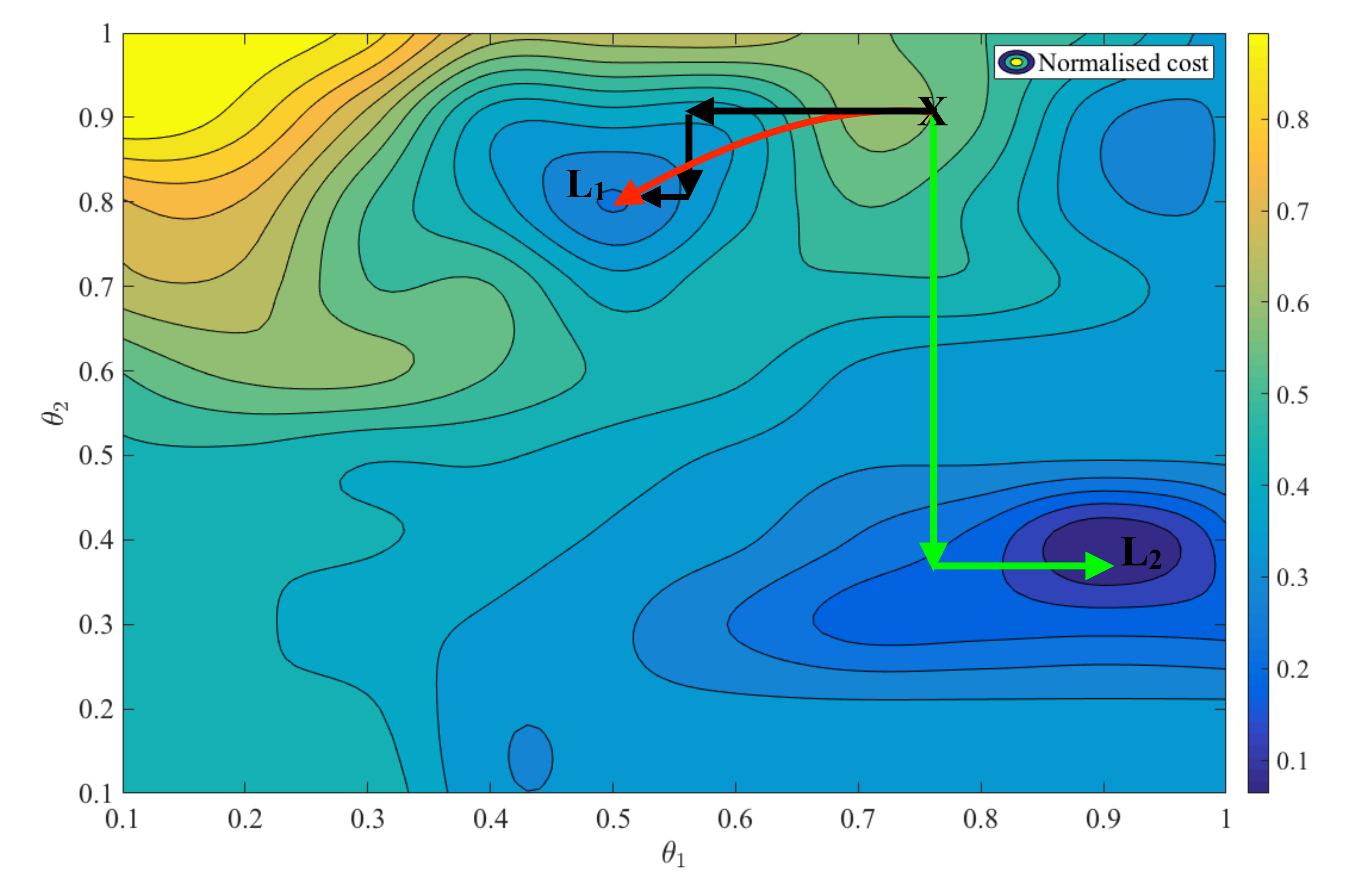}
\caption{Fictive second degree of freedom cost function. X denotes the starting point of the nonlinear optimisation. Depending on what parameters and the order in which the parameters are selected for optimisation, a different path along the cost function is followed. If both $\theta_1$ and $\theta_2$ are optimised simultaneously, in a gradient descent based way, the red path is followed, ending up in the poor local minimum $\text{L}_1$. If $\theta_1$ and $\theta_2$ are optimised alternatingly, starting by $\theta_1$, the optimisation would again end up in $\text{L}_1$ (black path). Starting by $\theta_2$, however, $\text{L}_1$ could be avoided, ending up in the more optimal minimum $\text{L}_2$ (green path).}
\label{f:costSchets_af}
\end{center}
\end{figure}

An example of a fictive, second degree of freedom cost function, is given in Fig.\ \ref{f:costSchets_af}. By optimising both parameters, $\theta_1$ and $\theta_2$, simultaneously, the red path is followed directly to the local minimum $\text{L}_1$. Alternatingly optimising both but starting from $\theta_1$, keeping $\theta_2$ constant to its initial value, results in the path indicated in black. Again ending up in $\text{L}_1$. Changing only the order of optimisation, now starting from $\theta_2$, creates the green path, which leads to the more optimal minimum $\text{L}_2$.

 In this work, the same strategy is applied to the optimisation of the PNLSS model. In that case, as will be explained in Section \ref{modelEst}, a total of 2772 parameters need to be optimised. Since the true optimal subset of parameters leading to a good minimum is unknown, a number of predefined subsets are evaluated. A typical subset would include all the linear coefficients in the $\textbf{A}, \textbf{B}, \textbf{C}$ and $\textbf{D}$ matrices and only a selection of the coefficients in $\textbf{E}$ and $\textbf{F}$. In total, we chose to evaluate 90 possible paths from the starting point. The possibilities are explained in detail in \ref{Appendix}. The one leading to the most optimal point is selected and the search is repeated. When none of the 90 options yields a better minimum, the optimisation is ended.
 A list of the tested subsets along with a schematic overview of the optimisation loop is given in Appendix A, Fig.\  \ref{f:OptScheme}.

\subsection{Estimating the initial state of the model}
\label{Estx0}
Given the fact that discrete PNLSS models are formulated recursively, $x(t+1) = f(x(t),u(t))$, the initial state $x_0$ and initial input $u_0$ need to be known at $t=0$, in order to be able to simulate the response of the model to a certain input sequence. To obtain values for $x_0$ and $u_0$, they can be seen as parameters that need to be estimated from the data. The problem is again nonlinear in the parameters and the Levenberg-Marquardt algorithm is used. As initialisation $x_0$ and $u_0$ are set to zero. At this stage, no guarantee of finding the true initial state and input can be given and errors can be introduced.

\review{When dealing with nonlinear systems, the impact of an error on $x_0$ and $u_0$ can be significant. This is due to the fact that there might exist multiple stable solutions (nonunicity) for a set of different initial conditions. Although a full sensitivity analysis revealing the bifurcation structure is out of the scope of this work, the impact of the initial state on the fluid force coefficient will be illustrated with an example in Section \ref{Estx0_cy}.}

\subsection{Validating the model}
\label{valModel}

How to assess the quality of the model will strongly \review{depend} upon the user. Generally, there exists a trade off between the accuracy of the model and the extent of parameter space covered. In this work we will demonstrate that a single model, valid over a wide range of parameter space, can be obtained at the expense of accuracy in some isolated regions. See Section \ref{model_val} for an in depth discussion on the validation.

\section{Application of the method}
\label{application}
In this section, the PNLSS identification procedure is applied to two systems: the nonlinear Van der Pol oscillator, and the problem of modelling \CYc. The Van der Pol oscillator is used as an illustration of the method since it can be shown that there exists an analytical PNLSS equivalent.

\subsection{The Van der Pol oscillator}
\label{VdP_example}
The Van der Pol oscillator is of special interest since such equations have been frequently used as wake oscillator models. One of their characteristics is that they are able to perform autonomous oscillations, and also describe a synchronisation behaviour when they are excited \cite{parkinson1974,hartlen1970}. The most basic representation is given by \cite{vanderpol1926}:
\begin{equation}
\label{e:wakeEq4}
\ddot{c}_y +\mu \Omega_{\textup{aut}} (c^2_y-1) \dot{c}_y + \Omega_{\textup{aut}}^2 \CY = \dot{y},
\end{equation}
where on the left hand side we have a nonlinear oscillator with an autonomous angular frequency  $\Omega_\textup{\small{aut}}$ and on the right hand side we have a forcing term related to the \review{derivative} of the displacement (forced Van der Pol equation). There is debate on what the best right hand side expression should be used to model VIV, displacement, velocity, or acceleration \cite{facchinetti2004}. In this article we use the velocity coupling. Here $\Omega_\textup{\small{aut}} = 2\pi f_{\textup{aut}}$ with $f_{\textup{aut}}$ the autonomous frequency. We chose $\mu$ equal to 0.3, in accordance with \cite{facchinetti2004}, to assure the characteristic behaviour of frequency synchronisation, limit cycle oscillations and autonomous oscillations are present.

\subsubsection{Visualisation of the lock-in range}
\begin{figure}[h!]
\begin{center}
\includegraphics[scale=0.25]{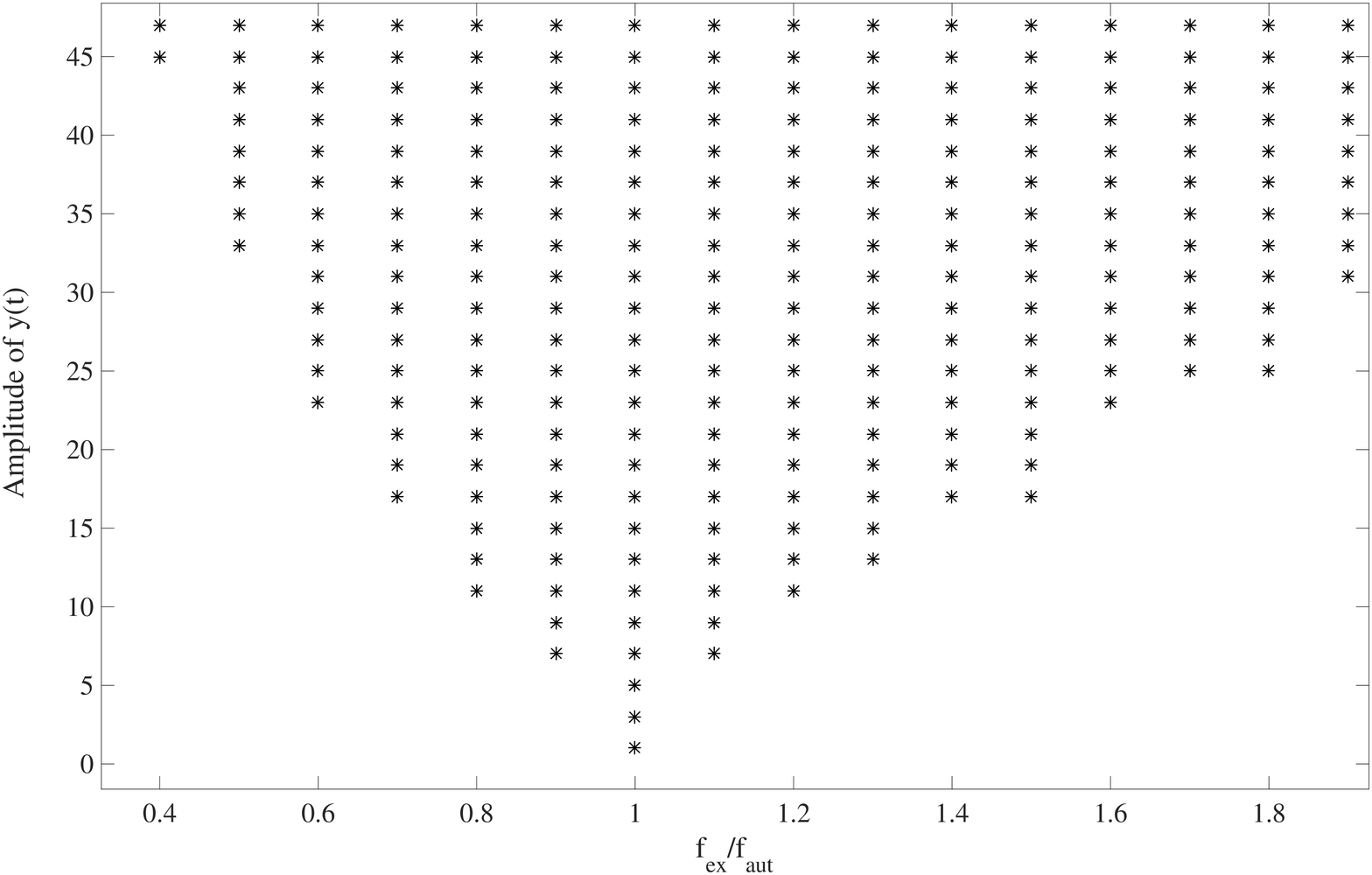}
\caption{Lock-in behaviour of the Van der Pol equation. Black stars mark the combinations of relative frequency and amplitude of y(t) for which the frequency of \CY in Eq.~\eqref{e:wakeEq4} is synchronised with the excitation frequency, $f_{\textup{ex}}$.}
\label{f:VdP_lock_in}
\end{center}
\end{figure}

We study the lock-in region by imposing a sinusoidal displacement, $y(t)= A\sin(2 \pi f_{\textup{ex}}t)$, of various amplitude levels and ranging in frequency around $f_{\textup{aut}}$. In Fig.\ \ref{f:VdP_lock_in}, the combinations of amplitude and relative frequency of $y$, for which the frequency of \CY locks on to the excitation frequency ($f_{\textup{ex}}$) are indicated with a black marker. A distinct v-shaped region, also called Arnold tongue \cite{pikovsky2001}, becomes visible. More information on the response and the bifurcation structure of the forced Van der Pol equation can be found in \cite{guckenheimer2003}.

\subsubsection{An analytical PNLSS expression of the Van der Pol equation}
It can easily be shown that the Van der Pol equation is covered by the PNLSS model class. Using a simple Euler discretisation, Eq.~\eqref{e:wakeEq4} can be rewritten in a discrete PNLSS form. With the state variables being $x(t) = [\CYc(t) \quad \dot{c}_y(t)]^\textup{T}$ and $T_s$ indicating the time step \review{the equivalent state-space formulation becomes:}
%
\begin{equation}
\label{e:PNLSS_VdP}
\begin{aligned} \left\{ \begin{matrix*}[l] \setlength{\arraycolsep}{6pt} x(t+1) &= \left[ \begin{matrix} 1 \quad T_s \\-T_s\Omega^2_{\textup{aut}} \quad 1+T_s\mu \Omega_{\textup{aut}} \end{matrix} \right] x(t) + \left[ \begin{matrix} 0 \\ T_s \end{matrix} \right] \dot{y}(t) + \left[\begin{matrix} 0 \\ \vspace{0.2cm} \review{-}T_s\mu \Omega_{\textup{aut}} \end{matrix}\right] c^2_y(t) \dot{c}_y(t) \\ \CYc(t) &= \left[ \begin{matrix} 1 \quad 0 \end{matrix} \right] x(t) \end{matrix*} \right. . \end{aligned}
\end{equation}
Since in the case of an oscillating cylinder in a fluid flow, the analytical expressions are unknown, we will also tackle the identification of the Van der Pol oscillator in a black-box way. We will show that a good estimate of Eq.~\eqref{e:PNLSS_VdP} can be found working solely from data, and thus without making use of any of the coefficients from the analytical expression, nor from the type (second order) of nonlinearity present. To get an estimate of the PNLSS model, steps 1 to 3 as described in Section \ref{IDPNLSS} are executed.

\subsubsection{Black-box PNLSS identification of the Van der Pol equation}
\begin{enumerate}
\item \textbf{Nonparametric estimate of the Best Linear Approximation}

By applying 7 different realisations of a random-phase multisine as input, $y$, in Eq.~\eqref{e:wakeEq4} and averaging the FRF over both the successive periods and the different phase realisations, nonparametric estimates $\hat{G}_{\text{BLA}}$ and $\hat{\sigma}_{BLA}$ are found. The multisines cover a frequency range from $f_0=0.1$ Hz to $4.5$ Hz ($1.5f_{\textup{aut}}$) with a flat amplitude spectrum \review{(in $y(t)$)}, $A_n=15$. A random-phase multisine realisation is given by Eq.~\eqref{e:MS1}, with $N=45$.The data are generated by solving Eq.~\eqref{e:wakeEq4} using the \emph{ODE45} built-in solver in MATLAB and a fixed time step of $10^{-3}$ s (small enough to limit integration errors). A time-domain and frequency-domain representation of the corresponding input and output can be seen in Fig.\ \ref{f:Input_output_VdP_vel}.

\begin{figure}[h!]
\begin{center}
\includegraphics[scale=0.25]{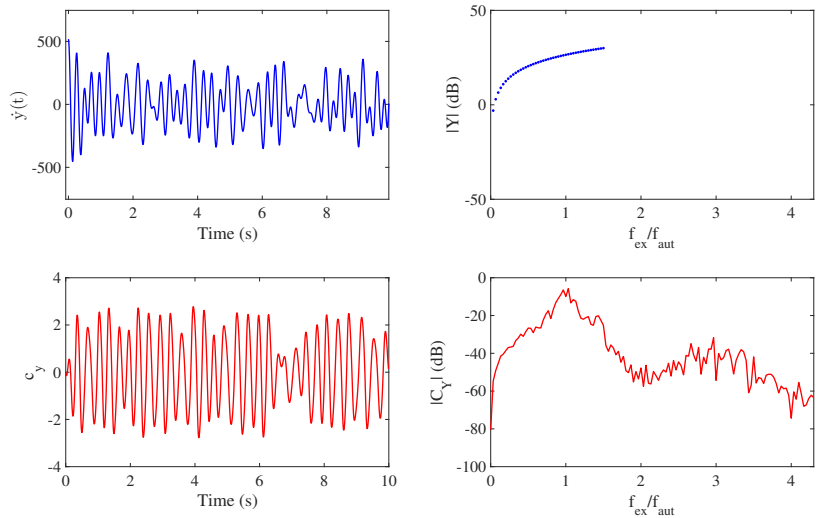}
\caption{Time and frequency domain plots of the input and output of the Van der Pol system. Top left: time series of $\dot{\textup{y}}(t)$. Bottom left: time series of \CYc. Top right: amplitude spectrum of $\dot{\textup{y}}(t)$ averaged over 5 periods. Since a flat amplitude spectrum was chosen for $y(t)$ ($A_n=15$), we have a linearly increasing spectrum in  $\dot{\textup{y}}(t)$. Bottom right: amplitude spectrum of \CY averaged over 5 periods.}
\label{f:Input_output_VdP_vel}
\end{center}
\end{figure}

\item \textbf{Parametric estimate of $\hat{G}_{\text{BLA}}$}

Using the FDIDENT toolbox in MATLAB, \review{a parametric model scan was conducted. In Fig.\ \ref{f:cost_VdP} the cost of the parametric fit is plotted against the model order. In this case a linear second-order model (1/2) was found adequate (clear drop in cost). Moreover it can be seen from the analytical expression in Eq.\ \ref{e:PNLSS_VdP} that 2 state variables would suffice.} 

\begin{figure}
\begin{center}
\includegraphics[width=\textwidth]{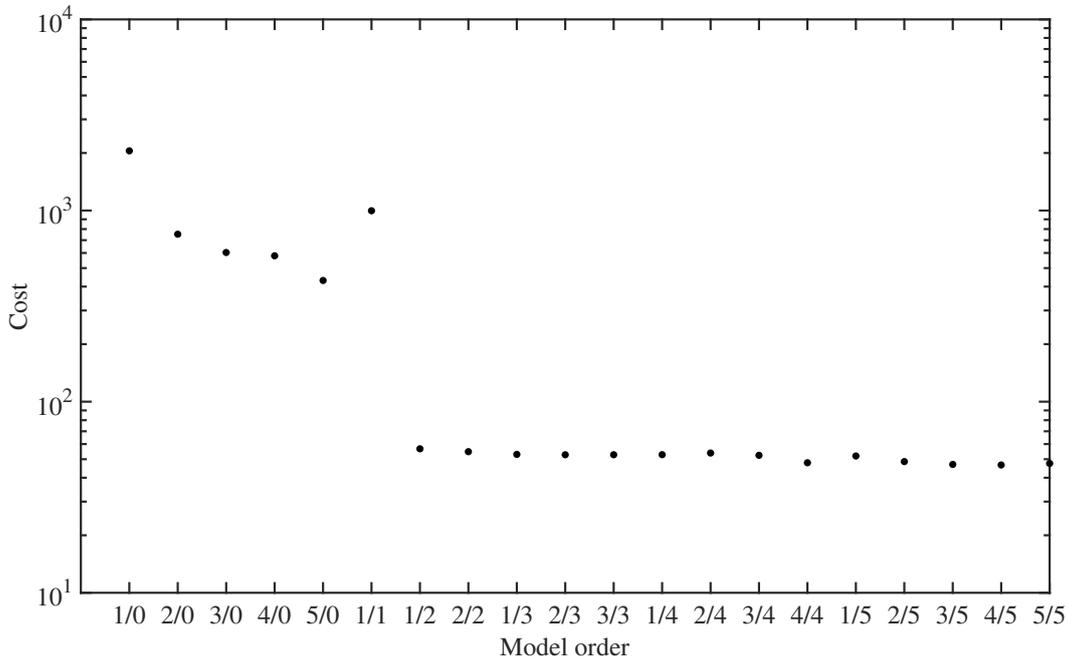}
\caption{Cost of the parametric fit of the Best Linear Approximation of the Van der Pol equation as a function of model order.}
\label{f:cost_VdP}
\end{center}
\end{figure}

A plot of the nonparametric and the parametric estimate of $\hat{G}_{\text{BLA}}$ can be seen in Fig.\ \ref{f:BLA_VdP}. \review{Also the level of total distortion and measurement noise are indicated in red and green correspondingly. Even though the data was not intentionally corrupted with noise, still a certain amount of distortions were measured. One major reason is that the autonomous frequency was not exactly equal to 3 Hz, because of integration errors when solving Eq.\ \ref{e:wakeEq4} in a discrete way. As a result leakage could not be avoided.}

\begin{figure}
\begin{center}
\includegraphics[scale=0.25]{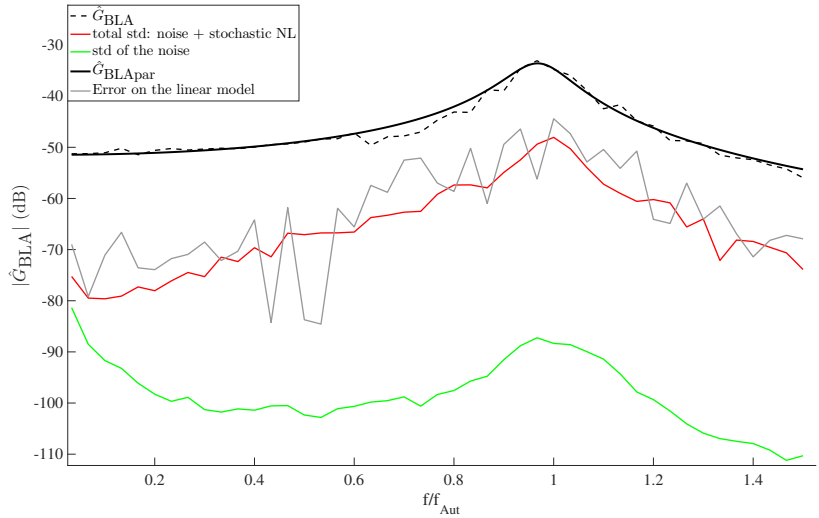}
\caption{Parametric (solid black line) and nonparametric (dashed black line) estimate of $\hat{G}_{\text{BLA}}$ of the Van der Pol system.}
\label{f:BLA_VdP}
\end{center}
\end{figure}

\item \textbf{Nonlinear optimisation}

Starting from $\hat{G}_{\text{BLA}}$ as initial values for the \textbf{A, B, C} and \textbf{D} matrices, the nonlinear optimisation strategy as explained in Section \ref{Optimisation_strategy} is applied to minimise the cost function:
\begin{equation}
\label{VLS_VdP}
V(\theta) = ||e(t,\theta)||^2 \quad \textrm{with} \quad e(t,\theta) = c_{y\text{,VdP}}(t)-c_{y\text{,mod}}(t,\theta),
\end{equation}
where $c_{y\text{,VdP}}$ is the data computed from Eq.~\eqref{e:wakeEq4} and $c_{y\text{,mod}}$ is the simulated output of the model.
The 7 multisine realisations are used as training data. \review{Since the objective is to model the Van der Pol equation in a black-box way, no precise selection of the nonlinear degrees was made. Instead, for both the state and the output equation all monomials ranging in degree from 2 up and till 5 were used (whereas judging from Eq.\ \ref{e:wakeEq4} a total degree 3 would be exact).} In the output equation also the DC term (degree 0) was included. \review{Theoretically, it should be possible to retrieve the exact model parameters (degree 3) by introducing the proper constraints on the \textbf{A, B, C} and \textbf{D} matrices. In practice, however, this is not straightforward, as there is no intuitive way to initialise the remaining coefficients in Eq.~\ref{e:PNLSS_VdP}.}
\end{enumerate}


After the linear identification step, the model showed an rms error of 33.1~\% when simulating the output of the estimation dataset. After running the nonlinear optimisation routine the relative rms error dropped to only 1.2~\%. Validation of this model on a new (unseen) phase realisation yields errors around 2.0~\% rms error. 

We have thus successfully retrieved a discrete PNLSS model that well describes the data obtained from Eq.~\eqref{e:wakeEq4} without imposing any prior knowledge on the PNLSS equations. Although an analytical PNLSS expression Eq.~\eqref{e:PNLSS_VdP} exists, we have to note that the exact coefficients are not retrieved (error does not drop to zero). This is the price to be payed for allowing for a larger flexibility in the nonlinear terms (degrees 0 till 5, while degree \review{3} would suffice), combined with the strong dependence on the initialisation of the linear part.

\subsection{Model estimation of \CY from CFD data}
\label{modelEst}
In this section, a model describing the kinematics of an oscillating cylinder in a fluid flow is estimated from CFD data.

\subsubsection{Initial values of the model: the Best Linear Approximation}
\label{BLA}
\begin{figure}[h!]
\begin{center}
\includegraphics[scale=0.25]{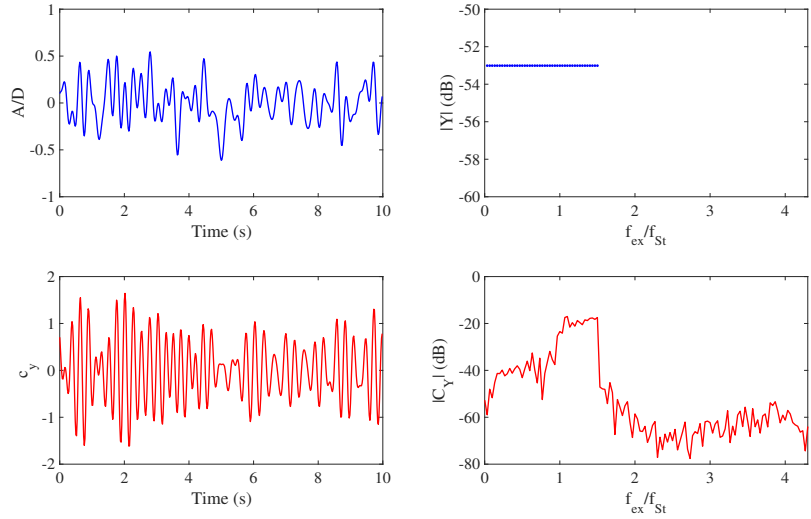}
\caption{Time and frequency domain plots of the input and output of the oscillating cylinder.Top left: time series of the displacement, $y$. Bottom left: time series of \CY. Top right: amplitude spectrum of $y$ averaged over 5 periods. Bottom right: amplitude spectrum of \CY~averaged over 5 periods.}
\label{f:input_output_Cy}
\end{center}
\end{figure}

As in the Van der Pol example in Section \ref{VdP_example}, the linear model is estimated using 7 random phase multisine realisations.~\review{From each realisation one period is regarded as transients and is not used for the identification.} In this case, contrary to the Van der Pol example where the input and output had no physical meaning, each realisation corresponds to a trajectory of an imposed cylinder oscillation. The frequency band covered ranges from 0 to $1.5f_{\text{St}}, (f_{\text{St}} = 3$ Hz), at a frequency resolution of $0.1$ Hz  and an amplitude level of $A/D=0.3$. In Fig.\ \ref{f:input_output_Cy}, one period of the input and the output, as \review{well} as their amplitude spectra, averaged over the successive periods, are plotted. The DFT of the input and the output are denoted $Y$ and $\BCYc$, respectively. 

\begin{figure}[h]
\begin{center}
\includegraphics[scale=0.35]{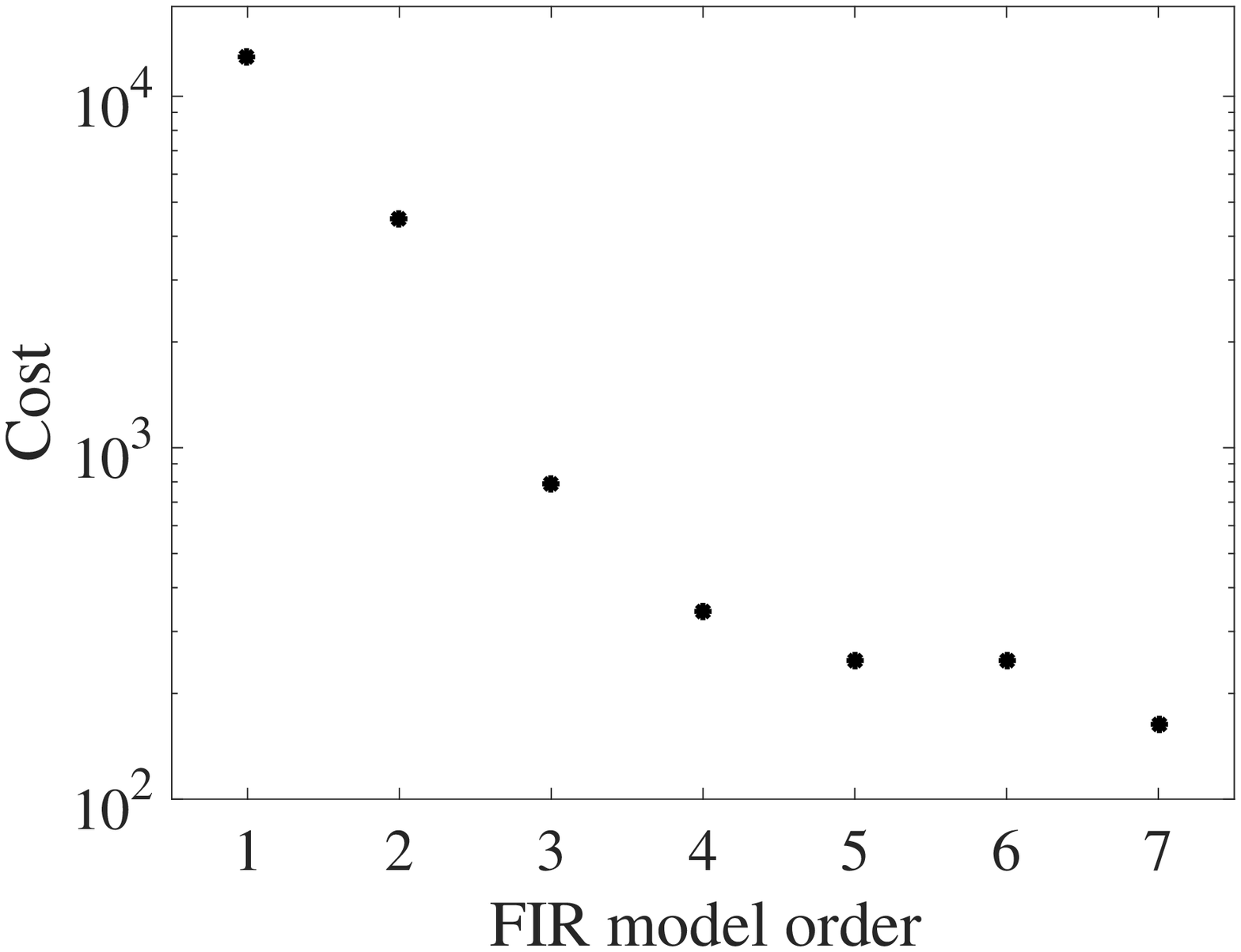}
\caption{Cost of the parametric estimate of $G_{\text{BLA}}$ as a function of the model order.}
\label{f:modelOrder_FIR}
\end{center}
\end{figure}

Given the nonparametric estimate ($\hat{G}_{\text{BLA}}$), found through application of Eq.~\eqref{e:average_spec} to (\ref{e:GBLA}), a parametric FIR (Finite Impulse Response) model was estimated. A FIR model was preferred to ensure stability \cite{oppenheim1998}. The \review{selection} of the model order is based on a trade off between the cost of the error of the fit (cost) and the number of state variables required. The higher the model order, the higher the number of state variables resulting in a combinatorial growth of monomials in $\zeta(t)$ and $\eta(t)$. A plot of the cost as a function of the model order is given in Fig.\ \ref{f:modelOrder_FIR}. An FIR model of 5th order was selected.

\begin{figure}
\begin{center}
\includegraphics[scale=0.25]{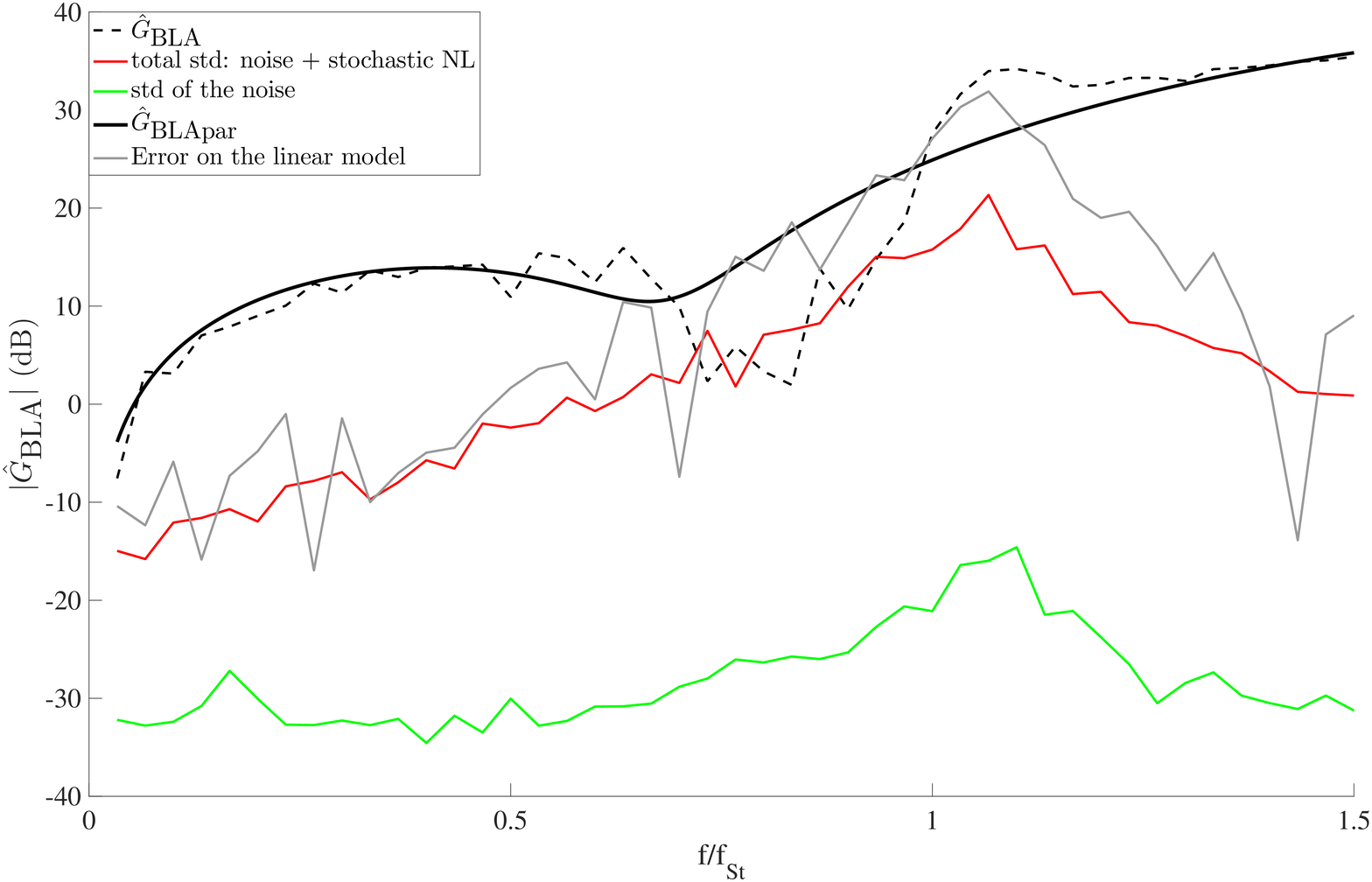}
\caption{The nonparametric and parametric estimates of the Best Linear Approximation of the transfer function relating \CY to $y(t)$ (dashed and solid black line respectively), computed from 7 different phase realisations of a multisine ranging from 0 to 4.5 Hz at $A/D=0.3$.}
\label{f:GBLA}
\end{center}
\end{figure}

Fig.\ \ref{f:GBLA} shows the nonparametric and the parametric estimate $\hat{G}_{\text{BLA}}$ and $\hat{G}_{\text{BLA,par}}$. The red line indicates the variation found over the different phase realisations that were applied. It is a measure for the combined effect of measurement noise and stochastic nonlinear influences (due to the random selection of the phases). Discrepancies between the nonparametric and the parametric estimates can be tolerated given the high uncertainty of the linear estimate due to nonlinear effects. The green line shows the noise floor. All components which are present in the signal but do not act at an integer multiple of $f_0$ will cause leakage. This will make up most of the power in what is considered to be noise. \review{From Eq.\ \ref{e:fSt} it follows that the Strouhal frequency is equal to 3 Hz. In practice we can however observe that this will not be exactly equal to 3. This will cause leakage to occur. Besides leakage also numerical averaging can cause fluctuations in the solution of the CFD simulations.} It can be seen that \review{especially in a band around $f_{\text{St}}$ the total level of distortion is of equivalent power as} the function itself, emphasising the need for a nonlinear model.
The parametric FIR estimate, written in state-space formulation, will serve as initial values for the \textbf{A}, \textbf{B}, \textbf{C} and \textbf{D} matrices in Eq.~\eqref{e:PNLSS2}.

\subsubsection{The estimation dataset}
\label{est_dataset}

The envisaged application of the identified model is the simulation of the forces on real-life systems such as an oil riser or a chimney stack. Generally, apart from gusts, the operation regime would consist of a constant (or slowly varying) flow speed resulting in a constant frequency of excitation and thus oscillation. Therefore, \review{accurate} simulation of the response to single sine oscillations will be most important.

When composing the training set, a trade-off must be made between signals that lean towards the intended operating condition of the model, which is a user requirement, and the compactness and richness of the data, a system identification requirement. To that end, sine-sweep excitations with a slow sweep rate were found most adequate. Including multisine excitations has also been tested but did not yield improved validation results on single sine validation tests, and for that reason they were omitted. The final training dataset was obtained by gradually increasing the amplitude density of the sine sweeps \review{in order} to improve the validation results. The sweeps cover the same frequency band as the multisines that were used during the linear identification step: 0 to $1.5f_{St}$ (0 - 4.5 Hz with a period $T_0=16$ s  for a single sweep direction) at amplitude levels from $A/D= 0.05$ to 0.3 in discrete steps of $0.025$. Table \ref{t:overview} gives an overview of the type of excitation signal and the identification technique used throughout the identification process.

\begin{table}[h!]
\begin{center}
\begin{tabular}{| l | c | c | c|}
\cline{2-4}
 \multicolumn{1}{c|}{} & model & excitation signal & identification technique\\
\hline
1 & nonparametric BLA & multisines & spectral averaging\\
2 & parametric BLA & multisines & FIR estimation\\
3 & PNLSS & swept sines & NL optimisation (LM)\\
\hline
4 & validation & single sines & \textbackslash \\
\hline
\end{tabular}
\end{center}
\caption{Overview of the identification steps and the corresponding excitation signals used.}
\label{t:overview}
\end{table}

\begin{figure}
\begin{center}
\includegraphics[width=\textwidth]{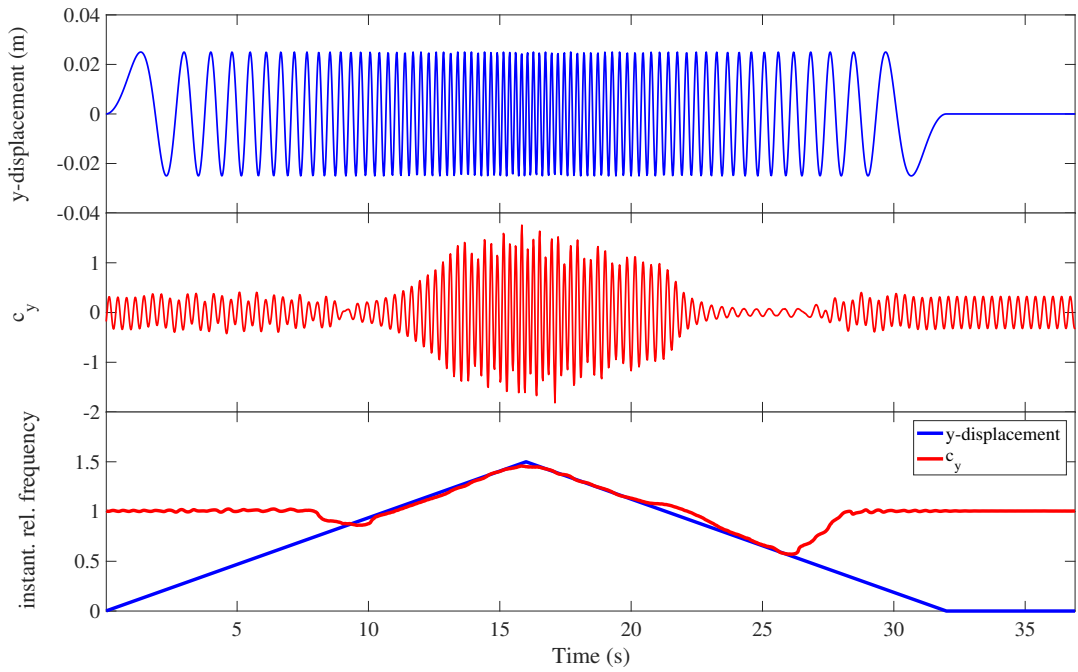}
\caption{The oscillation and the corresponding force Coefficient of an imposed sine sweep CFD simulation. Frequency ranges from 0 to $1.5f_{\text{St}}$, A/D = 0.25.}
\label{f:sweep_example}
\end{center}
\end{figure}

Fig.\ \ref{f:sweep_example} shows the imposed oscillation (on top) of one part of the training data ($A/D=0.25$). In the middle, the corresponding $\CY$is plotted and the bottom figure shows the relative instantaneous \review{frequency (relative to $f_{\text{St}}$) regarding both quantities}. Looking at the relative instantaneous frequency we can clearly observe synchronisation of the frequency of $\CYc$ between, roughly, 10 s and 26 s. Within this region, rapid variation in magnitude of $\CY$occurs. The different behaviour of the force coefficient during sweep up and down also highlights the hysteretic nature of the phenomenon. Oscillations in the relative instantaneous frequency are due to the presence of multiple frequency components in the signal. All sweep experiments end with a period in which the cylinder is held stationary so that the vortex shedding can relax to its autonomous behaviour. The stable limit cycle period is needed to ensure the smooth concatenation of different data parts in the training set (Fig.\ \ref{f:est_output}).

\subsubsection{Training of the model}
\label{training_model}
\begin{figure}
\begin{center}
\includegraphics[width=\textwidth]{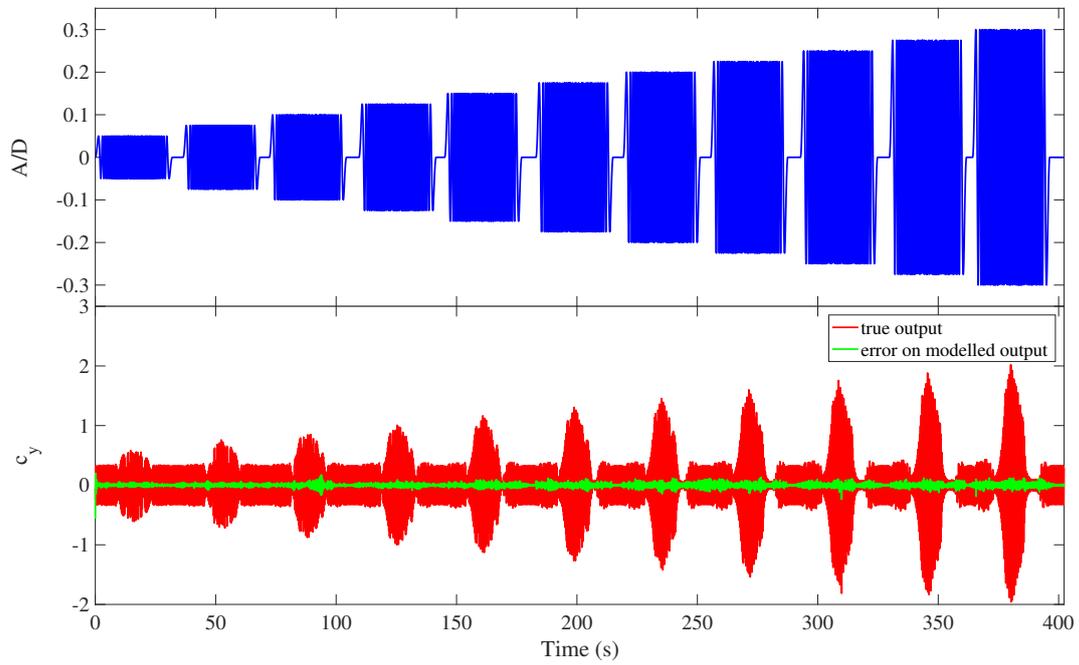}
\caption{The imposed oscillation during the training of the model (top figure), the true force coefficient obtained from CFD in red and the error on the modelled output in green (bottom figure).}
\label{f:est_output}
\end{center}
\end{figure}

Training of the model on the estimation data is done following the scheme in Fig.\ \ref{f:OptScheme} of Appendix A. Table \ref{t:est_results} gives an overview of how the error drops during successive iterations of the nonlinear optimisation over subsets of the coefficients. The selected subsets of the state and the output equation are indicated under `Optimisation choices'. Note that the iteration number in the left column refers to a cycle of the loop in Fig.\ \ref{f:OptScheme} and not to the underlying Levenberg-Marquardt algorithm which is executed in each iteration and typically ran for 1000 steps. The relative rms error, $e_{\text{rms}}$, between the modelled output and the true output of the estimation set is given by: 
\begin{equation}
\label{e:Erms}
\begin{multlined}
\begin{aligned}
e_{\text{rms}} = &\sum_{i=1}^n e_{\text{rms}_i} w_i,\\
e_{\text{rms}_i} = &\sqrt{\frac{\sum_{t = 0}^{t_{N_i}} (y_{\text{CFD}}(t)-y_{\text{mod}}(t))^2}{\sum_{t=0}^{t_{N_i}}y_{\text{CFD}}(t)^2}}, \\
w_i = &\frac{N_i}{N},
\end{aligned}
\end{multlined}
\end{equation}
where $N_i$ is the number of points of the $i^{th}$ data part and $N$ is the total number of points.
If, instead of iteratively optimising over subsets of the coefficients, all coefficients would have been tuned simultaneously, the optimisation would have hit a local minimum corresponding to a final error of $e_{\text{rms}}= 49.9$ \%, pointing out the relevance of the iterative approach. 
In Fig.\ \ref{f:est_output}, the input ($y$-displacement of the cylinder), output (force coefficient) and the error on the modelled output are plotted.

\begin{table}
\begin{center}
\begin{tabular}{ | c | l  | r |}
\hline
\textbf{iteration} & \textbf{Optimisation choices} & \textbf{$e_{\text{rms}}$} \\
\hline \hline
0 & Best Linear Approximation & 0.565 \\ \hline 
1 & $\begin{aligned} state\ equation:&\ \textup{affine in the states} \\ output\ equation:&\ \textup{All} \end{aligned}$ & 0.225 \\ \hline
2 & $\begin{aligned} state\ equation:&\ \textup{up to degree 3} \\ output\ equation:&\ \textup{affine in the states} \end{aligned}$ & 0.078 \\ \hline
3 & $\begin{aligned} state\ equation:&\ \textup{no cross products} \\ output\ equation:&\ \textup{All} \end{aligned}$ & 0.075 \\ \hline
\end{tabular}
\caption{List of the errors during training of the model on the estimation data.}
\label{t:est_results}
\end{center}
\end{table}

\subsubsection{Estimating the initial state of the model}
\label{Estx0_cy}
As described in Section \ref{Estx0}, the initial state and input, $x_0$ and $u_0$ also need to be estimated from the data. \review{It is known that nonlinear systems can have a non-unique solution following a bifurcation point. In \cite{blackburn1999} it was shown that the initial conditions, i.e.\ the starting time of the imposed motion with respect to the zero crossing of the autonomous oscillating force coefficient, can have a tremendous influence on the transient time before a steady state solution is obtained. The transients can range from only 15 motion cycles up to 200 cycles during which the solution changes continuously and unpredictably between cycles with positive and negative energy transfer. Moreover a complex higher order bifurcation structure was revealed while sweeping though the lock-in range at a constant amplitude level. In \cite{Elston2004,Elston2006} a stability analysis of the flow about imposed oscillating circular cylinders was conducted. It was found that different stable states existed, the distinction being based on the bifurcation that arises. One of the bifurcations is of the Neimark-Sacker type. This leads to quasi-periodic states.
But also outside lock-in it can be clear that the initial condition influences the quasi-periodic force coefficient. Since within this zone the wake will not synchronise to the cylinder motion, the starting time of the imposed motion will have a direct impact on the obtained solution. This is visualised by plotting the output to an imposed sweep and hold experiment for different initial conditions, Fig.\ \ref{f:x0mod}. The impact of errors on $x_0$ and $u_0$ will thus also be significant.} 

The test signal that is used is composed of two parts. In the first part the frequency sweeps up to $\rfrac{f_{\text{ex}}}{f_{\text{St}}}=0.5$, from then on the frequency is held constant. A high amplitude level of $\rfrac{A}{D}=0.25$ is applied. In the top part of Fig.\ \ref{f:x0mod}, the output of the model for two different $x_0$ values is plotted. The bottom part shows the instantaneous frequency of the input. $u_0$ is set to zero since at all time steps prior to the start of the input, the cylinder was indeed stationary. Clearly, the influence of the initial state remains present throughout the entire signal length. \review{We want to stress that although it cannot be guaranteed that the exact bifurcation structure of the underlying system is also present in the model, this influence of the initial state is clearly a system property and not due to numerics.}
\begin{figure}
\label{f:x0mod}
\begin{center}
\includegraphics[scale=0.25]{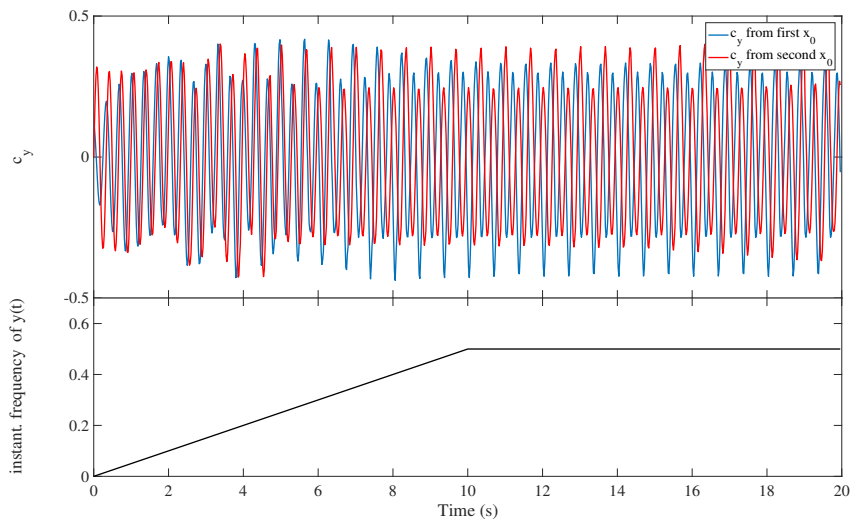}
\caption{Top figure: simulation output \CY of the model, given two different initial states on an identical input signal. Bottom figure: instantaneous frequency of the input signal, y(t).}
\label{f:x0mod}
\end{center}
\end{figure}
\subsubsection{Model validation} 
\label{model_val}

To study the validity of the estimated model, it is subjected to various validation experiments. How to validate the model strongly depends on the \review{intended} application, and is therefore a user choice. In this work, two user requirements are put forward:
\begin{enumerate}
\item We want the model to perform well on signals which relate closely to how VIV is observed in practice. Therefore the validation experiments will be of the type of single sine oscillations.
\item As a second requirement, we want a single model to be valid over a wide range in frequency and amplitude of the input signal. The set of validation experiments will thus consist of a large number single sine oscillations of different frequencies and amplitudes.
\end{enumerate}

As quality measures, four quantities are computed for each validation experiment:
\begin{itemize}
\item The time domain relative rms error, $e_{\text{rms}}$ Eq.~\eqref{e:Erms}
\item The correlation coefficient $R$
\item The relative error on the maximum of the amplitude,
\begin{equation}
 e_{\text{maxA}} = \frac{c_{y\text{max}}-c_{y\text{max}}(\theta)}{c_{y\text{max}}},
\end{equation}
This is an interesting measure since the maximum amplitude is a strong design criterium.
\item We also want to introduce a quality measure which is phase independent, since we have seen that the model is very sensitive to errors on the estimate of the initial state (Section \ref{Estx0_cy}) and since we know that such errors will mainly manifest themselves as phase errors. This is obtained by transforming the error to the frequency domain. A frequency-domain error, based only on the magnitudes of the DFT of the true output, $C_Y$, and the simulated output, $C_{Y_{\text{mod}}}$ is given by:
\begin{equation}
e_{\text{DFT}} = \sqrt{\frac{\sum_{\omega=0}^{\omega_s}(|C_Y|-|C_{Y_{\text{mod}}}|)^2 \Delta(\omega)}{\sum_{\omega=0}^{\omega_s}|C_Y|^2\Delta(\omega)}},
\end{equation}
where $\omega_s=2\pi f_s, f_s$ being the sampling frequency.
\end{itemize}
An overview of the applied validation experiments and the corresponding errors is given in Table \ref{t:val_mono}. 
\newpage
\begin{center}
\begin{longtable}[!ht]{ | c | c | c | c | c | c |}
\cline{2-6}
 \multicolumn{1}{c|}{} & $\begin{aligned} \textup{\textbf{Excitation}} \\  \rfrac{f_{\text{ex}}}{f_{\text{St}}} \quad A/D \end{aligned}$  & \textbf{$e_{\text{rms}}$} & \textbf{R} & \textbf{$e_{\text{maxA}}$} & \textbf{$e_{\text{DFT}}$} \\
\cline{2-6}
\hline
0 & 0 \quad 0 & 0.03 & 0.99 & 0.004 & 0.03\\
\hline
1 & 0.5 \quad 0.05 & 0.11 & 0.99 & 0.014 & 0.10\\
2 & 0.5 \quad 0.10 & 0.23 & 0.97 & 0.049 & 0.23\\
3 & 0.5 \quad 0.15 & 0.09 & 0.99 & 0.057 & 0.05\\
4 & 0.5 \quad 0.20 & 0.19 & 0.98 &  0.024 & 0.08\\
5 & 0.5 \quad 0.25 & 0.43 & 0.89 & 0.041 & 0.23\\
\hline
6 & 0.7 \quad 0.05 & 0.06 & 0.99 & 0.017 & 0.05\\
7 & 0.7 \quad 0.10 & 0.10 & 0.99 & 0.032 & 0.07\\
8 & 0.7 \quad 0.15 & 0.17 & 0.98 & 0.059 & 0.16\\
9 & 0.7 \quad 0.20 & 0.26 & 0.96 & 0.006 & 0.22\\
10 & 0.7 \quad 0.25 & 0.22 & 0.97 & 0.086 & 0.18\\
\hline
11 & 0.9 \quad 0.05 & 0.19 & 0.98 & 0.014 & 0.16\\
12 & 0.9 \quad 0.10 & 0.38 & 0.93 & 0.020 & 0.11 \\
13 & 0.9 \quad 0.15 & 0.19 & 0.98 & 0.000 & 0.11 \\
14 & 0.9 \quad 0.20 & 0.14 & 0.99 & 0.028 & 0.12\\
15 & 0.9 \quad 0.25 & 0.12 & 0.99 & 0.007 & 0.09\\
\hline
16 & 1.1 \quad 0.05 & 0.26 & 0.96 & 0.033 & 0.25\\
17 & 1.1 \quad 0.10 & 0.51 & 0.86 & 0.046 & 0.27\\
18 & 1.1 \quad 0.15 & 0.66 & 0.75 & 0.047 & 0.15\\
19 & 1.1 \quad 0.20 & 0.27 & 0.97 & 0.116 & 0.13\\
20 & 1.1 \quad 0.25 & 0.05 & 0.99 & 0.032 & 0.03\\
\hline
21 & 1.3 \quad 0.05 & 0.05 & 0.99 & 0.006 & 0.04 \\
22 & 1.3 \quad 0.10 & 0.08 & 0.99 & 0.022 & 0.07\\
23 & 1.3 \quad 0.15 & 0.22 & 0.97 & 0.020 & 0.04\\
24 & 1.3 \quad 0.20 & 0.07 & 0.99 & 0.044 & 0.05\\
25 & 1.3 \quad 0.25 & 0.17 & 0.98 & 0.007 & 0.07\\
\hline
\cline{2-6}
\multicolumn{1}{c|}{} & \textbf{mean} & 0.20 & 0.96 & 0.032 & 0.12\\
\multicolumn{1}{c|}{} & \textbf{std} & 0.15 & 0.05 &  0.026 & 0.07\\
\cline{2-6}
\caption{Validation results of mono sine oscillations.}
\label{t:val_mono}
\end{longtable}
\end{center}

Each validation experiment starts with a sweep up, until the desired frequency is reached. For the remaining part of time the frequency is held constant. The quality measures are calculated only on the \review{part where the desired frequency is reached and held constant}.

\begin{figure}[h!]
\begin{center}
\begin{subfigure}[b]{\textwidth}
\includegraphics[width=\textwidth]{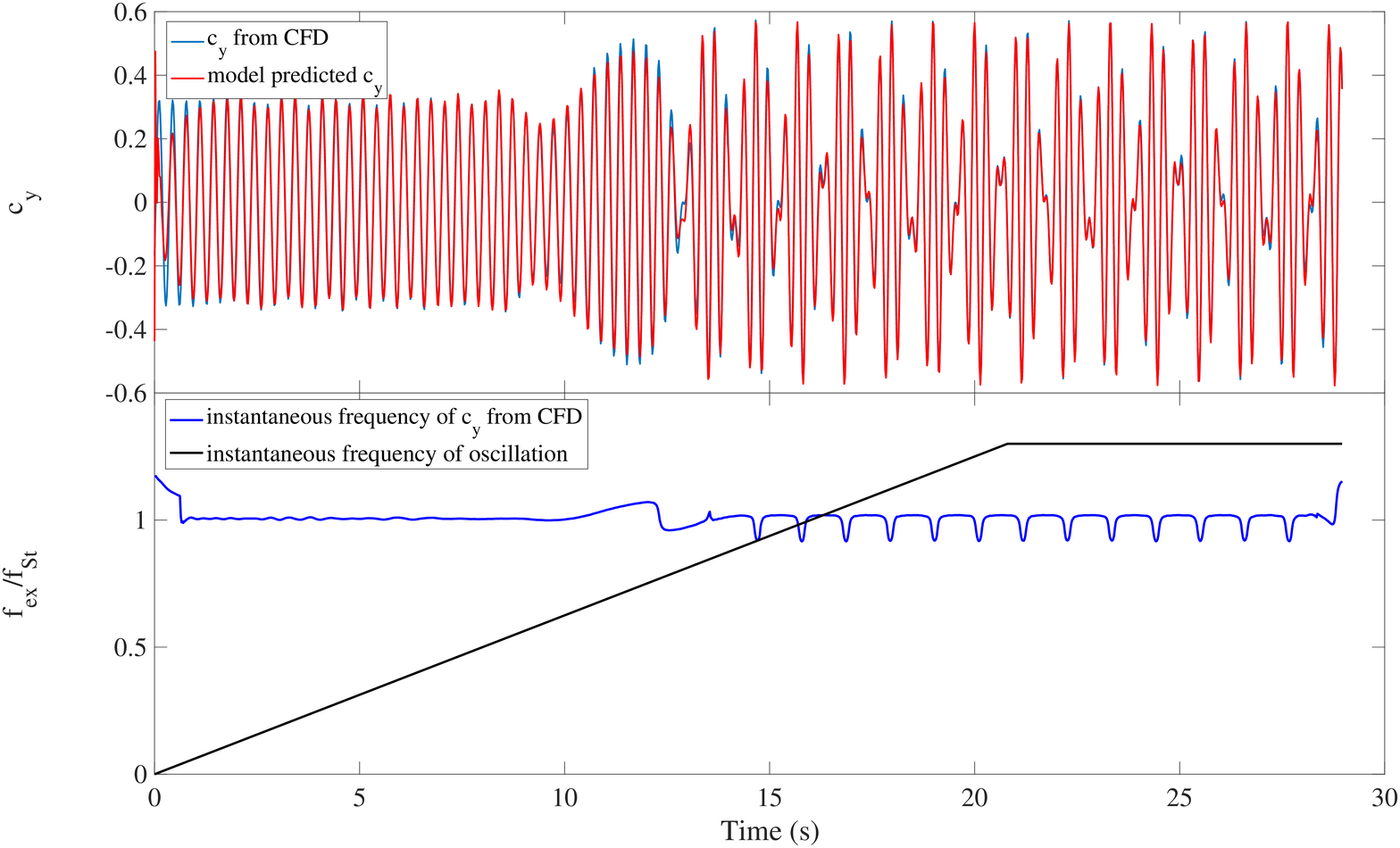}
\caption{}
\label{f:val_mono_nice}
\end{subfigure}
\begin{subfigure}[b]{\textwidth}
\includegraphics[width=\textwidth]{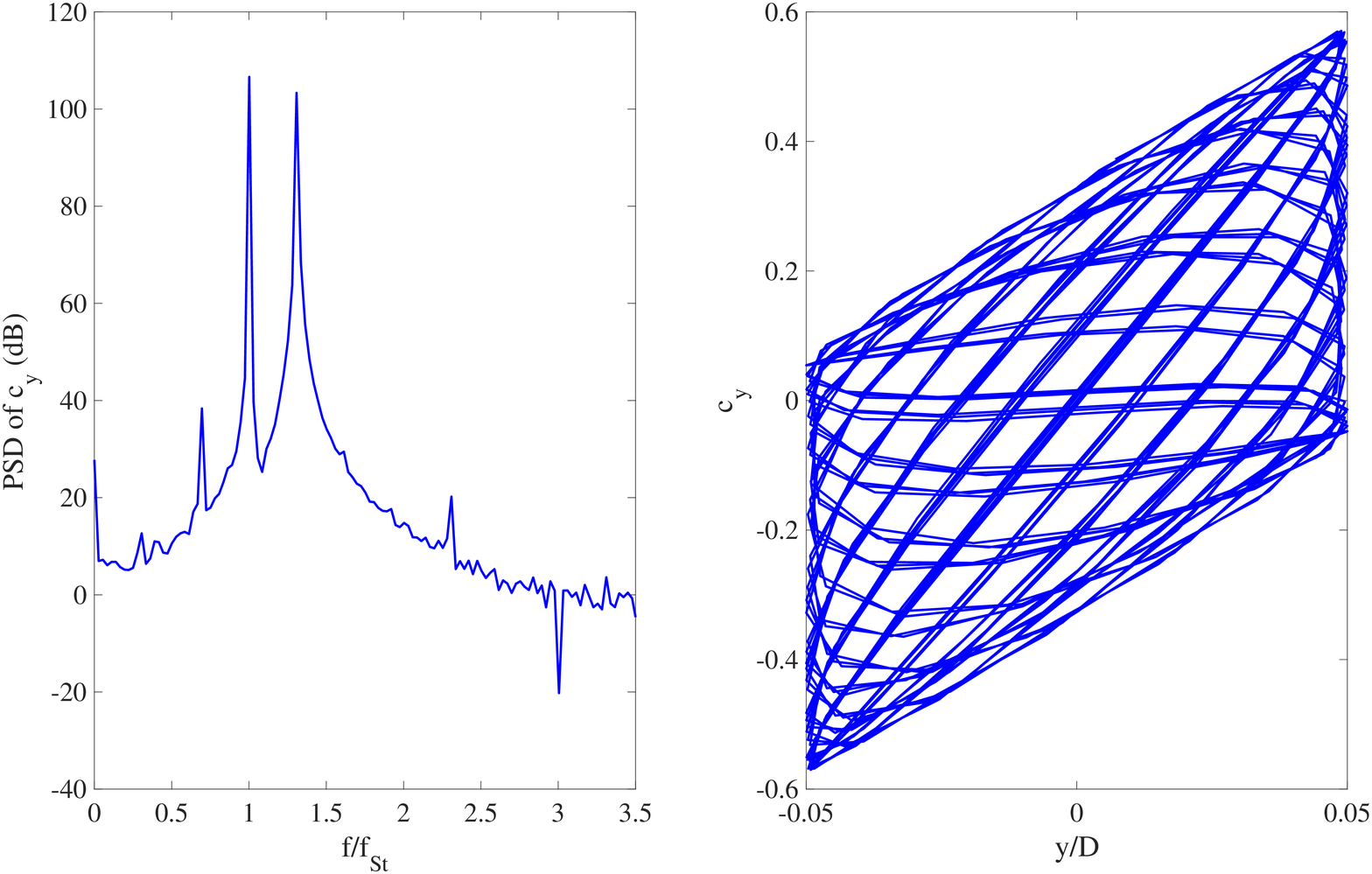}
\caption{}
\label{f:val_mono_nice_2}
\end{subfigure}
\caption{Figure (a): time domain output for mono sine validation experiment 21. In the bottom part the instantaneous frequency (relative to $f_{\text{St}}$) is plotted. Figure (b): PSD of \CY (from CFD) in the left figure, phase-plot of \CY versus the relative displacement $y/D$ indicating quasi-periodicity on the right.}
\end{center}
\end{figure}

By plotting the true output and the simulated output together as time series it was noticed that in practically all cases the model was able to accurately reproduce the maximum amplitude, the correct frequency and even the characteristic features of the signal. In Fig. \ref{f:val_mono_nice}, the simulated output of validation experiment 21 is plotted. Although the case is outside of lock-in, a strong interaction between the natural vortex shedding and the imposed oscillation is visible. The quasi-periodicity becomes even more clear when looking at the PSD and the phase plot of \CY (Fig.\ \ref{f:val_mono_nice_2}). The model is able to reproduce this quasi-periodic oscillation with an accuracy of 95 \%.

\begin{figure}[h!]
\begin{center}
\begin{subfigure}[b]{\textwidth}
\includegraphics[width=\textwidth]{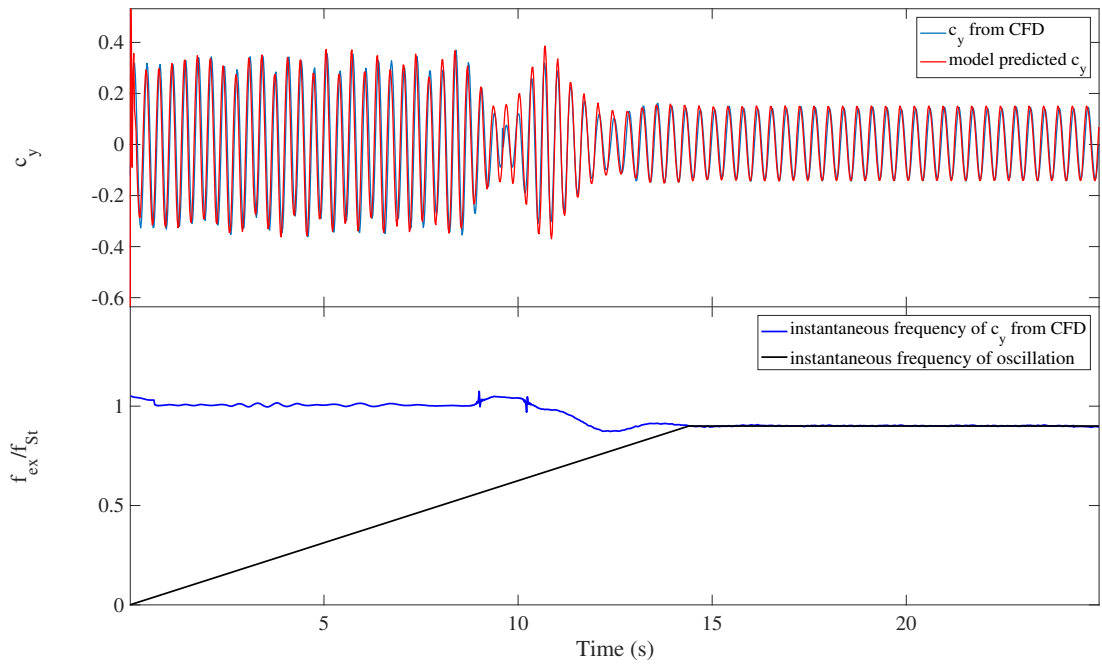}
\caption{}
\label{f:val_mono1}
\end{subfigure}
\begin{subfigure}[b]{\textwidth}
\includegraphics[width=\textwidth]{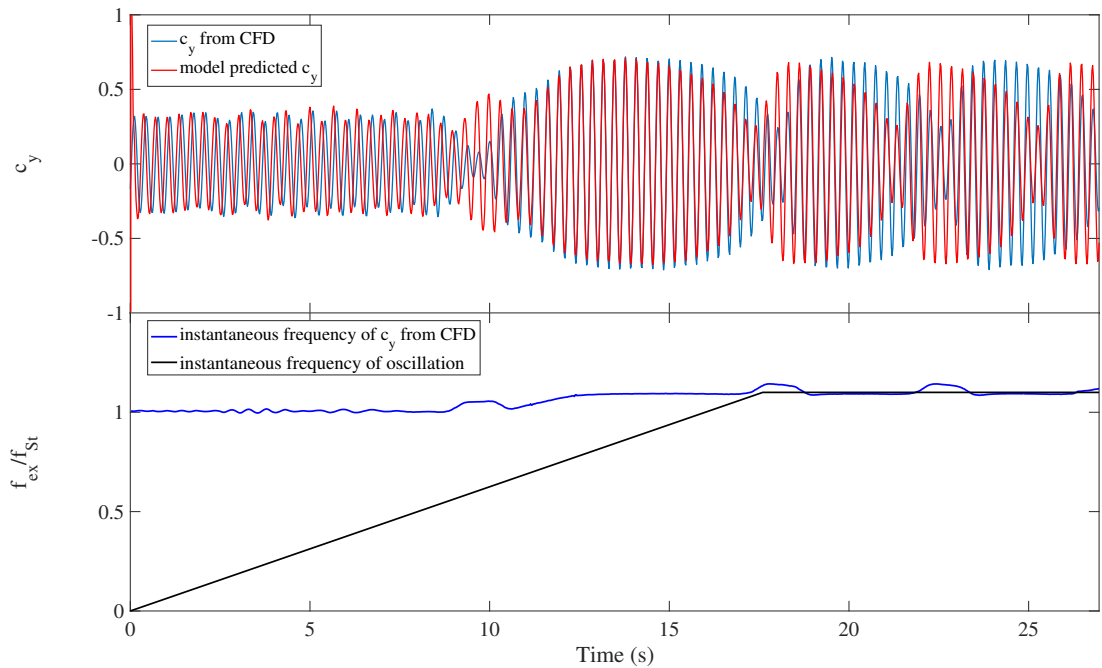}
\caption{}
\label{f:val_mono2}
\end{subfigure}
\caption{Figure (a): time domain output for mono sine validation experiment 12. Figure (b): time domain output for mono sine validation experiment 17. In the bottom parts of both figures the instantaneous frequency (relative to $f_{\text{St}}$) is plotted \review{in order} to visualise wether or not \CY is synchronised to the excitation.}
\end{center}
\end{figure}

In a few cases, higher values of $e_{\text{rms}}$ were observed. This increased error appears to be due to small phase errors between the true and the simulated output, for which $e_{\text{rms}}$ is a very sensitive measure (an example of such case suffering from a phase error is given in Fig.\ \ref{f:val_mono1}). Phase errors are easily introduced due to the sensitivity on the initial state (see Section \ref{Estx0_cy}). As a result we can have two solutions of the model, which track the same limit cycle, although they will never be equal to each other at a single time instant.

From a user point of view, having small phase errors on the output will in many cases be tolerable, given that a good estimate of the \review{maxium} amplitude, the frequency of the response and the feature of the signal are obtained. In these cases, the quality of the simulated output can best be quantified in terms of $e_{\text{DFT}}$.

In Fig.\ \ref{f:val_mono2}, the time domain output of the case which performs the worst in terms of $e_{\text{DFT}}$ is plotted (experiment 17). The oscillating behaviour of the instantaneous frequency of \CY measured from CFD suggests that this case resides on the border of lock-in, exhibiting quasi-periodic oscillations \review{(see Section \ref{Estx0_cy})}.\review{This region of parameter space is also known as the transition region (transition to lock-in) \cite{kumar2016}.} Although this very sensitive instability is also captured by the model, the entering and exiting of synchronisation is out of phase with respect to the true output. Altering the initial conditions of the CFD \review{set up}, i.e.\ the point at which the cylinder oscillation starts interacting with the natural vortex shedding, revealed that the fluid is also able to stabilise to multiple slightly different solutions. This highlights the sensitivity of this particular part of parameter space.

\begin{figure}[h!]
\begin{center}
\begin{subfigure}[b]{\textwidth}
\includegraphics[width=\textwidth]{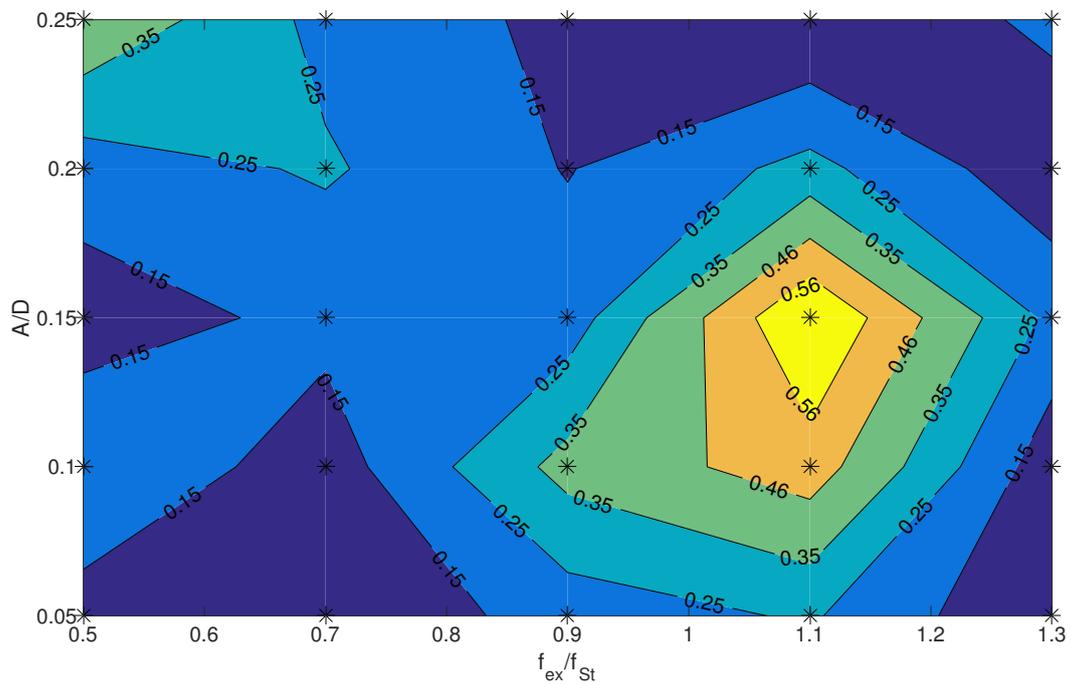}
\caption{}
\label{f:contour_relrms_shifted}
\end{subfigure}
\begin{subfigure}[b]{\textwidth}
\includegraphics[width=\textwidth]{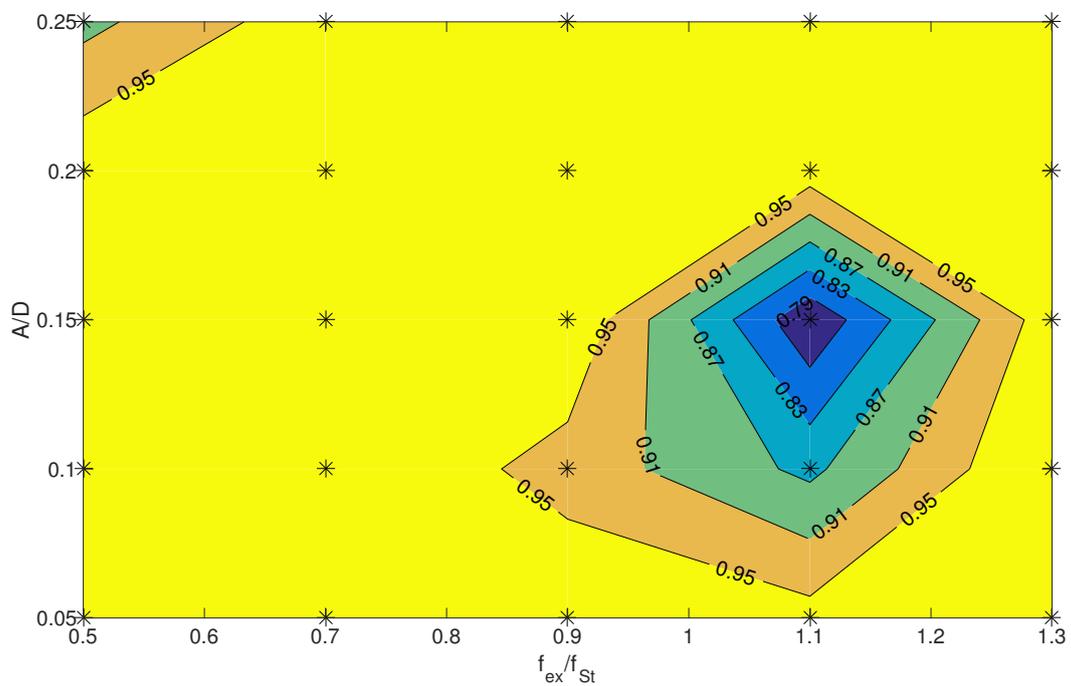}
\caption{}
\label{f:contour_corr_shifted}
\end{subfigure}
\end{center}
\caption{In Figure (a) a contour plot of the relative rms error is presented in the frequency-amplitude plane. Figure (b) shows a contour plot of the correlation coefficient. In both figures, the locations of the validation experiments are indicated with black markers, the colours are interpolated between these markers.}
\end{figure}

Since the correlation coefficient $R$ has a mean value of 0.96 with only a limited spread we can conclude that the modelled output is in good relation with the true output. The estimated model is, in other words, able to reproduce the distinct behaviour of the force coefficient both inside and outside the lock-in region for cylinders oscillating at a single frequency. A contour plot of the relative rms error (Fig.\ \ref{f:contour_relrms_shifted}) and of $R$ (Fig.\ \ref{f:contour_corr_shifted}) visualises the results in the frequency-amplitude plane.

\begin{figure}[h!]
\begin{center}
\begin{subfigure}[b]{\textwidth}
\includegraphics[width=\textwidth]{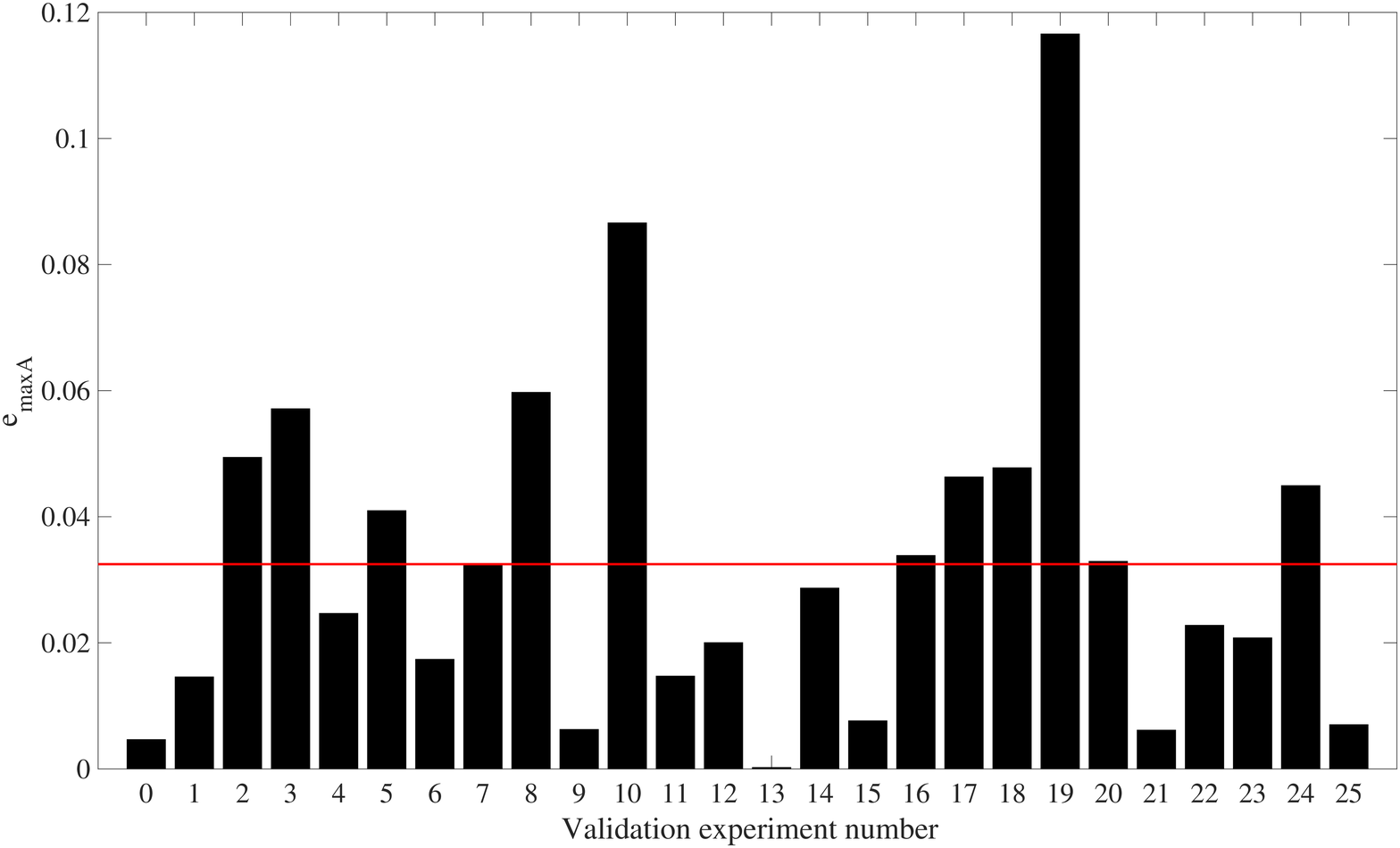}
\caption{}
\label{f:MaxAmpErrorA}
\end{subfigure}
\begin{subfigure}[b]{\textwidth}
\includegraphics[width=\textwidth]{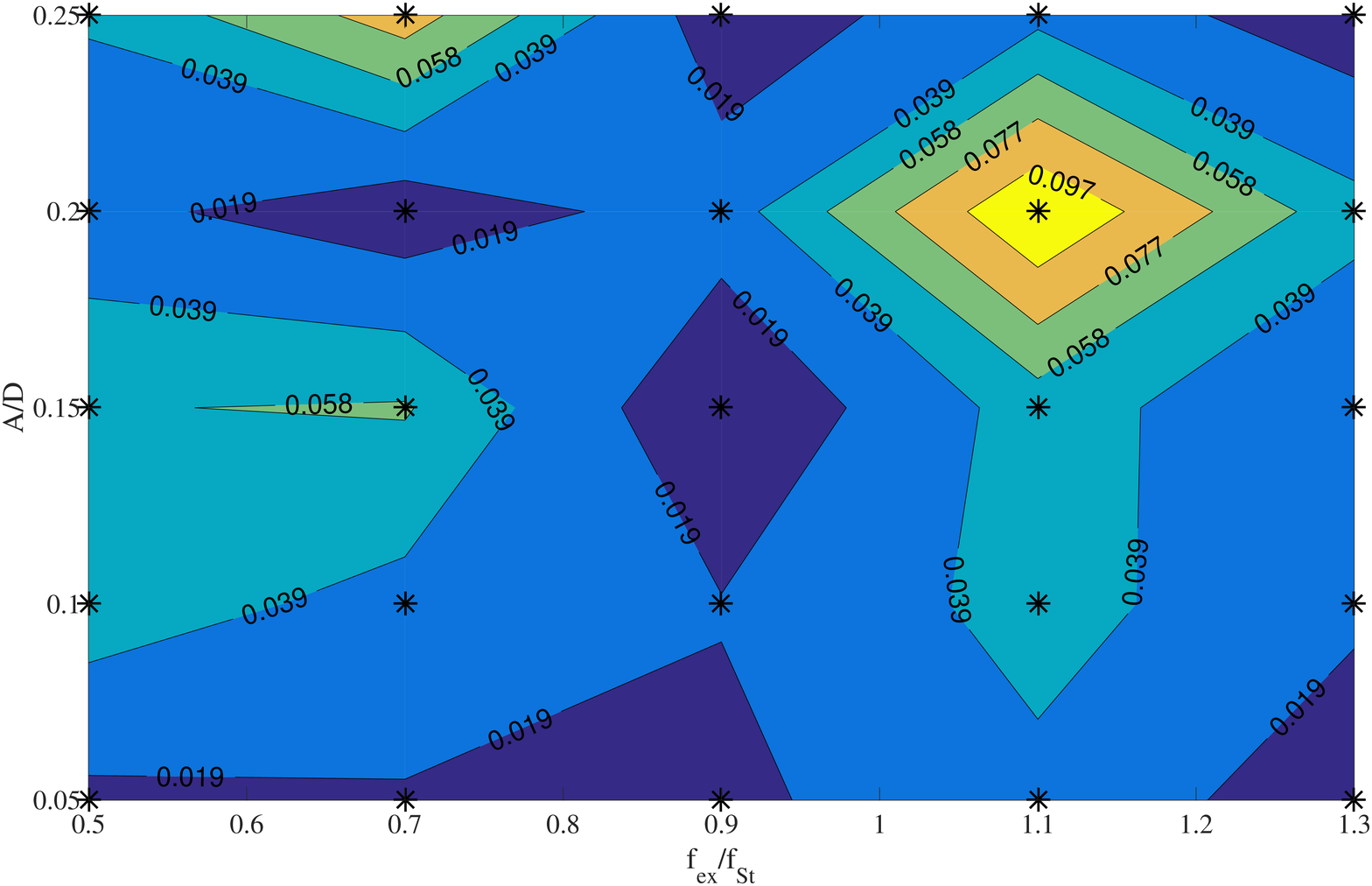}
\caption{}
\label{f:MaxAmpErrorB}
\end{subfigure}
\end{center}
\caption{Figure (a): The relative error on the maximum amplitude of all validation experiments. The red line indicates the mean value of 3.2~\%. Figure (b): Contour plot of the relative error on the maximum amplitude in the frequency-amplitude plane.}
\label{f:MaxAmpError}
\end{figure}

Considering potential applications of the model, e.g.\ for design purposes, the relative error on the maximum of the amplitude might well be the most crucial quality measure. A low mean value of 3.2~\% and only a small spread can be reported here (Fig.\ \ref{f:MaxAmpError}). 

\section{Conclusions}
\label{conclusion}
In this work, a discrete PNLSS model that relates the displacement of an oscillating cylinder in a fluid flow to the fluid forces acting on the cylinder is identified. The model has been trained on a concatenation of imposed swept sine oscillations and it has been validated on single sine experiments, as single sines are considered to be the intended application of the model. Data, both for training \review{and} for validation were obtained using CFD simulations at the fixed free-stream condition corresponding to a Reynolds number of 100. The PNLSS model is initialised using the best linear approximation, obtained from exciting the system with random phase multisine realisations. Subsequently, an iterative optimisation strategy is used to avoid poor local minima. The identification approach has numerically been validated on data obtained from the Van der Pol equation.
The quality of the fluid-force model was assessed based on: the relative rms error, the relative error on the maximum amplitude, the correlation coefficient and a phase independent error. It was shown that the output of the model was in good agreement with the CFD data over a wide range of oscillation frequencies and amplitudes. Sensitive regions in the frequency-amplitude plane, especially on the edge of the lock-in region, were observed to be hard to simulate in terms of phase. The feature and the maximum amplitude, however, remained accurate.

\section*{Acknowledgements}
This work was supported in part by the Fund for Scientific Research (FWO-Vlaanderen), by the Flemish Government (Methusalem), the Belgian Government through the Inter-university Poles of Attraction (IAP VII) Program, and by the ERC advanced grant SNLSID, under contract 320378.


\appendix
\setcounter{figure}{0}
\section{Graphical representation of the iterative optimisation scheme}
Iterative optimisation over subsets of the total set of coefficients have shown to contribute in avoiding local minima of the cost functions. The iterations can be schematically represented by a loop (Fig.\ \ref{f:OptScheme}). By evaluating multiple possible subsets during each iteration of the optimisation loop, different paths along the cost function are tracked. Doing so, a path around a local minimum can be discovered. Since the optimal subset, \review{which would lead} to a path on the cost function that avoids a local minimum is unknown, a number of predefined subsets are tested during each iteration of the optimisation loop. The proposed subsets are: the total $\textbf{A}, \textbf{B}, \textbf{C}$ and $\textbf{D}$ matrices and
\begin{itemize}
\setlength\itemsep{0 em}
\item  The diagonal elements of \textbf{E}. All monomials where a single state is raised to a power up to $p$ while all other powers equal zero.
\item All monomials with only contributions in the input.
\item All monomials that are affine in the states. Monomials where only one state appears, and it appears linearly.
\item All monomials without cross products.
\item All monomials that are full state-affine. Monomials where at most one state appears, and if it appears, it appears linearly.
\item All monomials with total degree up to 3.
\item All monomials with only contributions in the states.
\item Only odd nonlinear degrees.
\item All monomials except for the DC component (all states as well as the input raised to the power 0).
\item All monomials.
\end{itemize}

Clearly, also other subsets could have been proposed. Given the fact that 10 possible subsets were proposed (of which the diagonal option is only valid for the state equation, the $\textbf{E}$ matrix) and not necessarily the same choice has to be made for the state as well as for the output equation, we arrive at a total of 90 possibilities (paths on the cost function) to be tested during each optimisation run.

\begin{landscape}
\label{Appendix}
\begin{figure}[h!]
\centerline{
\includegraphics[scale=0.6]{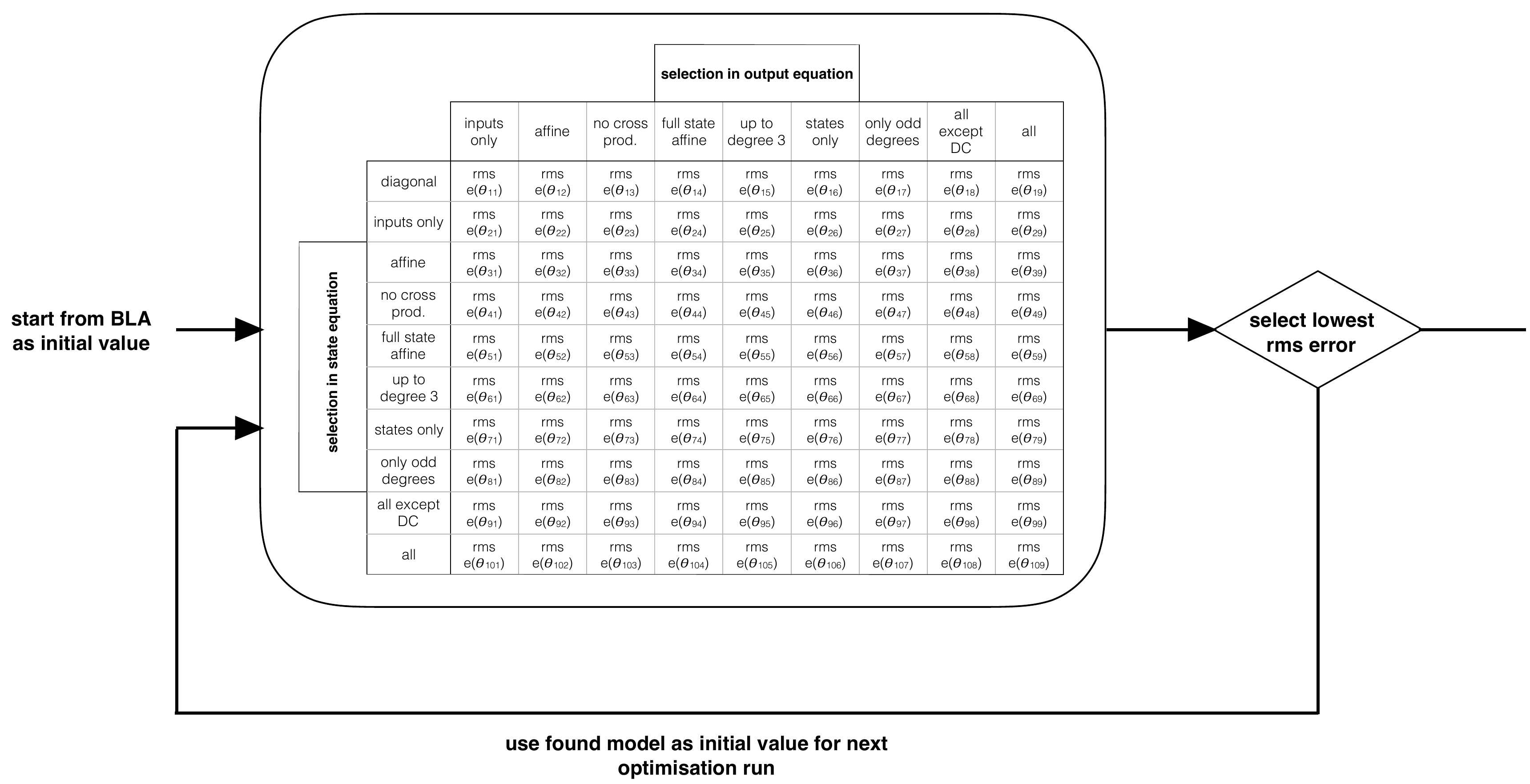}}
\caption{Optimisation scheme. The centre block represents the step in which the optimisation is carried out for all the possible combinations of subsets of parameters for the state and the output equation. Every combination yields a different rms error value.}
\label{f:OptScheme}
\end{figure}
\end{landscape}





\bibliographystyle{elsarticle-num} 
\bibliography{aerobib}

\end{document}